\documentstyle[11pt,aaspp4]{article}

\slugcomment{To be published in The Astrophysical Journal, 2001, 546, Jan 10 issue}

\lefthead{Gonz\'alez Delgado et al.}
\righthead{Seyferts 2}

\begin{document}

\title{The Nuclear and Circum-nuclear Stellar Population in Seyfert 2 Galaxies:
 Implications for the Starburst-AGN Connection}

\author{Rosa M. Gonz\'alez Delgado}
\affil{Instituto de Astrof\'\i sica de Andaluc\'\i a (CSIC), Apdo. 3004, 18080 Granada, Spain}
\affil{Electronic mail: rosa@iaa.es}

\author{Timothy Heckman}
\affil{Department of Physics \& Astronomy, JHU, Baltimore, MD 21218}
\affil{Adjunct Astronomer at STScI}
\affil{Electronic mail: heckman@pha.jhu.edu}

\and

\author{Claus Leitherer}
\affil{Space Telescope Science Institute, 3700 San Martin Drive, Baltimore, MD
21218}
\affil{Electronic mail: leitherer@stsci.edu}


\newpage

\begin{abstract}

We report the results of a spectroscopic investigation of a sample of 20
of the brightest type 2 Seyfert nuclei. Our goal is to search for the
direct spectroscopic signature of massive stars, and thereby probe the
role of circumnuclear starbursts in the Seyfert phenomenon. The method used 
is based on the detection of the higher order Balmer lines and 
HeI lines in absorption and the Wolf-Rayet feature at $\sim$4680 \AA\ in emission. 
These lines are strong indicators of the presence of young (a few Myrs) and 
intermediate-age (a few 100 Myrs) stellar populations. In over half
the sample, we have detected HeI and/or strong stellar absorption 
features in the high-order (near-UV) Balmer series together with relatively weak lines
from an old stellar population. In three others we detect a broad emission feature 
near 4680 \AA\ that is most plausibly ascribed to a population of Wolf-Rayet stars 
(the evolved descendants of the most massive stars). We therefore conclude that the 
blue and near-UV light of over half of the sample is dominated by young and/or intermediate age
stars. The ``young'' Seyfert 2's have have larger far-IR luminosities,
cooler mid/far-IR colors, and smaller [OIII]/H$\beta$ flux ratios than
the ``old'' ones. These differences are consistent with a starburst playing
a significant energetic role in the former class. We consider the possibility
that there may be two distinct sub-classes of Seyfert 2 nuclei (``starbursts''
and ``hidden BLR''). However, the fact that hidden BLRs have been found in
three of the ``young'' nuclei argues against this, and suggests that
nuclear starbursts may be a more general part of the Seyfert phenomenon.  
 
\end{abstract}



\newpage
\section{Introduction}

According to the standard unified scheme for radio-quiet AGN, 
Seyfert 2 (S2) galaxies contain a Seyfert 1 (S1) nucleus that is obscured by 
a circum-nuclear torus of dust and gas. However, electrons and/or dust grains 
located along the polar axis of the torus can reflect the light from the hidden 
nucleus, allowing it to be seen in polarized light. Generally 
speaking the optical continuum of S2 nuclei can be classified into two components:
a red continuum which is very similar to that of elliptical 
galaxies or early-type galactic bulges, and a bluer continuum in which the stellar lines 
are inconspicuous (the `featureless continuum', or FC).
According to the unified picture, the FC was believed to be scattered light from the 
hidden S1 nucleus. It dominates the ultraviolet continuum and contributes
between 10$\%$ and 50$\%$ to the optical light. 

However, this interpretation is not completely correct because after the 
contribution of the old stars is removed, the remaining optical continuum has a 
significantly lower fractional polarization than the broad optical emission lines 
(Tran 1995a,b,c). 
Therefore another unpolarized component has to contribute to the FC. 
Tran (1995c) suggests that the unpolarized component is optically thin thermal 
emission from warm gas, that is heated by the central hidden source. In contrast, 
Cid-Fernandes \& Terlevich (1995) and Heckman et al. (1995) suggest that the 
unpolarized component is a young but dust-reddened stellar population, possibly associated with 
the obscuring torus. 

Evidence that the unpolarized FC component in S2 nuclei is associated with young massive stars 
has been found in the last two years. In fact, to investigate the origin of the FC in 
S2 galaxies and the possible connection between
the starburst and Seyfert phenomenon, we started a program to study the ultraviolet (UV) 
continuum emitted by their nuclei with the goal of detecting stellar features from
massive stars. We choose this spectral range because it is the most favorable
wavelength range to detect massive stars, since these stars emit most of their flux at UV and far-UV
wavelengths. In contrast, older stars contribute significantly at optical and longer
wavelengths. Therefore, UV images provide potentially direct information about the location of
recent unobscured star forming regions. UV images of
starburst galaxies show that a significant fraction of the massive stars are formed in
very compact stellar clusters (Meurer et al. 1995) that have sizes of a few pc. Therefore, 
the morphology of the UV light in S2 nuclei can indicate whether these nuclei harbor a starburst. 
However, conclusive evidence for the presence of a starburst in the nuclei of S2 galaxies
can be obtained from their UV spectrum. The UV
spectrum of a starburst is very rich in absorption lines due both to stars and
interstellar gas. The strongest stellar lines are formed in the wind of massive
stars. These lines are blueshifted by 2000-3000 km s$^{-1}$ and/or show a P-Cygni profile.
Stellar photospheric lines are also present, but these features are much weaker than the
wind resonance lines and are not shifted in velocity. Strong resonance absorption lines also
form in the interstellar gas. These lines are often blueshifted by several hundred km
s$^{-1}$ (Gonz\'alez Delgado et al. 1998a). However, these interstellar lines cannot be taken 
as a proof of the existence of a starburst, because a non-stellar central engine can also
drive a similar kind of outflow. Conclusive proof that the UV emission in the S2 nuclei 
comes from a starburst is obtained if stellar wind resonance lines 
(such as NV $\lambda$1240, SiIV $\lambda$1400 and CIV $\lambda$1550) are detected in their spectra. 

These stellar wind resonance lines were detected with Hubble Space Telescope's
 (HST) Goddard 
High Resolution Spectrometer (GHRS) in four (Mrk 477, 
NGC 7130, NGC 5135 and IC 3639) of the brightest UV S2 nuclei of our sample (see section 2). The data 
indicate that the UV light is dominated by the starburst component (Heckman et al. 1997, 
Gonz\'alez Delgado et al. 1998b; hereafter H97 and GD98, respectively), with the contribution of the
hidden nucleus amounting to less than 20\%\  of the total emission. Comparing the strength of the
wind lines  with those predicted by evolutionary synthesis models (Leitherer et al. 1995), we
conclude that in these S2 nuclei, the starbursts are of short duration, with ages between 4 and 6
Myr. We find that, after  correcting the observed UV continuum for dust extinction using the
starburst attenuation law (Calzetti, Kinney, \& Storchi-Bergmann, 1994), the bolometric luminosity of these starbursts
($\sim$ 10$^{10}$ L$\odot$) is similar to the estimated luminosity of the hidden S1 nucleus. 
Signatures of massive stars have been found also in other active galactic nuclei (AGN). Maoz et
al. (1998) detect stellar features from massive stars in three of the seven brightest UV LINERs
observed with HST. These nuclei seem to belong to the LINER sub-class that Heckman (1996) and
Filippenko \& Terlevich (1992) have proposed may be metal-rich starbursts. In these objects
the estimated bolometric luminosity of the starbursts ranges between
$\sim$ 10$^8$ to 10$^9$
L$\odot$. This is significantly lower than that in S2 nuclei, but is similar to the
luminosities of super-star clusters that are believed to be the building blocks
of starburst galaxies. Therefore, we conclude from this previous work that, at least in the UV-
brightest S2 and LINERs, the light is often dominated by a starburst. Hot young stars are the origin of the
unpolarized FC component in at least some S2 nuclei. 

Additional evidence for the presence of massive stars has been found in the optical spectra 
of a few Seyfert
galaxies. At 4680 \AA\, the spectra of these Seyfert nuclei show a broad emission
feature (H97; Storchi-Bergmann, Cid Fernandes \& Schmitt 1998 (S98); Tran et al. 1999;
Kunth \& Contini 1999). The origin of this broad emission cannot be due to scattered light
from the hidden S1 nucleus, because their polarized spectra do not show this feature. Alternatively,
it has been proposed that this emission has a stellar origin, produced by Wolf-Rayet
stars, the descendants of massive stars. However, there is no
additional evidence for the presence of a starburst in the nuclei of these Seyferts, except in
Mrk 477. In this S2 nucleus, the detection of ultraviolet resonance lines formed in the wind of
massive stars strongly supports the interpretation that the broad emission at 4680~\AA\
is produced by Wolf-Rayet stars (H97). 

Another approach to study the origin of the FC has been carried out by Cid Fernandes and
collaborators (Cid Fernandes, Storchi-Bergmann \& Schmitt 1998, hereafter CF98). They have measured
the optical continuum flux in selected windows and the equivalent widths of several optical stellar
absorption lines (mainly CaII K and H, the G band, MgI+MgH feature and the NaI D doublet) as a
function of the distance from the nucleus for  a sample of Seyfert galaxies. The main result
obtained is that the stellar lines toward the nucleus  are not significantly diluted, suggesting
that if a FC is present, it contributes less than 10$\%$ to the optical light. The exception
to this behaviour is found in objects that harbor a  nuclear starburst. In these cases, the
equivalent widths of the stellar lines are strongly diluted.  Spectral synthesis using the stellar
lines listed above indicate that the Seyfert 2 nuclei of  their sample have a larger contribution
from stars with ages of 100 Myr (Schmitt, Storchi-Bergmann \& Cid Fernandes 1999).

These previous results indicate that at least the brightest UV S2 nuclei harbor
a starburst. Their UV images suggest that these starbursts have sizes that are similar to
those of the Narrow Line Regions (a few 100 pc). However, these results do not demonstrate
that starbursts are an ubiquitous part of the Seyfert phenomenon. Cid Fernandes \&
Terlevich (1995) suggested that circum-nuclear starbursts are a direct consequence of the
existence of a molecular torus in Seyfert galaxies, and thus a fundamental part of the Seyfert
nuclei. In fact, the torus is the natural place where the star formation can be triggered
because it contains a reservoir of molecular gas at a radius of the order of 100 pc. 

To more deeply investigate the origin of the FC in S2 nuclei and the role that starbursts play
in the Seyfert phenomenon, we present near-ultraviolet to near-infrared  spectra of a sample of
the 20 brightest S2 nuclei readily accessible from Kitt Peak, with the goal of detecting stellar
features  characteristic of
starbursts. Here, we describe a method to detect young and intermediate massive stars, based 
on the detection of Balmer and HeI photospheric lines. The paper is organized as follows. In
Section 2  we present the criteria used to select the objects. Section 3 presents the
observations and data reduction. Section 4 describes the method followed to detect starburst
features at optical wavelengths. Results and interpretation of the data are in Sections 5-8. 
We discuss our results and their implications in Section 9, and summarize our conclusions in
Section 10.

\section{Sample selection}

The sample comprises the brightest Seyfert 2 nuclei selected from the compilation of Whittle
(1992a,b).  They have been selected by their nuclear [OIII] $\lambda$4959,5007 emission line flux 
and by their non-thermal radio continuum emission at 1.4 GHz. They all satisfy
at least one of the two following criteria: log F$\rm_{[OIII]}\geq$ -12.0 (erg cm$^{-2}$ s$^{-1}$)
and  log F$_{1.4}\geq$-15.0 (erg cm$^{-2}$ s$^{-1}$). The criteria used imply 
that the sample should be unbiased with respect to the presence or absence of a nuclear
starburst. All the objects are classified as spiral galaxies, with 2/3 of the galaxies 
having a morphological type earlier than Sc (the remaining galaxies do not have a more specific
determination)\footnote{The morphological type is from NED.}. The redshifts range from 1000 km s$^{-1}$ to
15000 km s$^{-1}$,  with a mean value of 6450 km s$^{-1}$, giving a mean spatial scale of 420 pc
arcsec$^{-1}$ ($H_0$ = 75 km s$^{-1}$ Mpc$^{-1}$). 
At least ten galaxies in the sample are morphologically peculiar. Of these, two galaxies
(Mrk 273, Mrk 463E)
may be the result of mergers, seven (Mrk 1, Mrk 348, Mrk~477, NGC~5929, NGC 7212 and
 possibly NGC 7130 and IC 3639)
are in gravitational interaction with close companions, and one, Mrk~533, belongs to a Hickson Compact Group
and shows signs of tidal interaction. Another three galaxies (NGC 1068, NGC 5135 and Mrk 1066) are members
of a groups. The targets and their general properties  are summarized in Table 1.

\section{Observations and data reduction}

The observations were carried out using the Richey-Chr\'etien spectrograph attached to the 4m
Mayall telescope at Kitt Peak National Observatory during two runs in 1996 February and October. 
The detector was a T2KB CCD chip with a spatial
sampling of 0.7 arcsec/pixel. The slit width was set to 1.5 arcsec for all the observations 
and it was oriented near the parallactic angle, with the exception of NGC 1068 for which it was set at
P.A.=123$^{\circ}$.
 We used two gratings, KPC-007 and BL420, to cover 
the blue and the red spectral ranges with a dispersion of 1.39 \AA/pixel and 1.52 \AA/pixel, 
respectively. Since the edges of the chip are vignetted by the spectrograph camera, the useful
spectral ranges are $\sim$ 3400-5500 \AA\ (for the blue) and $\sim$ 6600-9100 \AA\ or $\sim$
7400-9800 \AA\  (for the red). Typical exposure times were 2$\times$1200 s and 2$\times$1800 s for
the blue and red  spectral ranges, respectively. The journal of observations is in Table 2.

The reduction of the spectra was performed with the FIGARO data processing software. Bias
subtraction and flatfielding were performed in the usual way; a two
dimensional wavelength calibration was done using ARC2D (Wilkins \& Axon
1991). The spectra were corrected for atmospheric extinction using a
curve appropriate for the observatory. The frames were flux calibrated
using the standards observed
during the same nights with the same set up, but with a 10 arcsec slit
width. The frames were corrected for distortion along the spatial
direction by using a second order polynomial fit to the spectral
distribution of the peak continuum emission of the standard stars, and
then applied to the galaxy frames. Sky subtraction was performed in
each data frame using the outermost spatial increments where no
obvious emission from the galaxy was present. Finally, frames of the same
wavelength range were co-added.

In addition to the targets of the sample, we observed the nuclei of three galaxies:
NGC 1569 (a dwarf irregular galaxy), NGC 205 (a dwarf elliptical galaxy), and NGC 4339
(an elliptical galaxy). The nuclear spectra of these galaxies show characteristics of
young ($\leq$ 10 Myr), intermediate (a few hundred Myr), and old (10-15 Gyr)
stellar populations, respectively. Analysis of the spectra of NGC 1569 is presented in Gonz\'alez
Delgado et al. (1997), of NGC 205 in Gonz\'alez Delgado et al. (1999) and Bica, Alloin \& Schmidt
(1990), and of NGC 4339 in  Ho, Filippenko \& Sargent (1995). 
The spectra of these objects are used as templates for the analysis of the Seyfert galaxies.

\section{Method for detecting starburst characteristics}

Our UV observations of a few objects (those with the brightest UV emission) have shown
direct spectroscopic detection of a starburst in the nuclei of these S2 galaxies using
strong stellar wind lines like NV $\lambda$1240,  SiIV $\lambda$1400 and CIV
$\lambda$1550. However, we do not know whether these few galaxies belong to a particular
class of Seyfert 2 nuclei or, instead, if starbursts are a  fundamental part of the
Seyfert phenomenon. To establish the possible connection between the starburst and
Seyfert phenomenon we need to detect starburst characteristics in a larger sample of S2
nuclei. Unfortunately, UV spectroscopic observations of  this kind of objects are
difficult to obtain because only a few of these nuclei have high enough UV surface brightness
to be  observed spectroscopically with HST.  In contrast, near-ultraviolet to
near-infrared observations of these objects are more feasible from ground based
facilities. Therefore, other diagnostics for starburst characteristics
in Seyfert galaxies at optical wavelengths are needed. 
To search for other diagnostics, first, we
examine the  morphology of the optical continuum of classical nuclear starbursts. 

The optical continuum of classical starburst galaxies is dominated by O, B and A stars. 
The spectra of young and intermediate mass stars are characterized by strong hydrogen
Balmer and neutral  helium absorption lines, and with only weak metallic lines (Walborn
\& Fitzpatrick 1990). The detection of these stellar features at optical wavelengths is
difficult because H and HeI absorption features  are coincident with strong nebular
emission-lines that mask them. However, the higher-order  terms of
the Balmer series and some of the HeI lines are detected in absorption in many starburst
galaxies (Storchi-Bergmann, Kinney, \& Challis, 1995)  and in the spectra of giant  HII
regions (e.g. NGC 604, Gonz\'alez Delgado \& P\'erez 2000). These features can be seen
in absorption  because the strength of the Balmer series in emission decreases rapidly
with decreasing  wavelength, whereas the equivalent width of the stellar absorption
lines is constant  or increases with wavelength (Gonz\'alez Delgado, Leitherer \&
Heckman 1999).  Thus, the net effect is that H$\alpha$, H$\beta$ and H$\gamma$ are
mainly seen in emission (but they can also show absorption wings superposed on the
nebular emission), but the higher-order terms of the series are seen mainly in
absorption. A similar effect occurs for the HeI lines, with HeI $\lambda$3919, HeI
$\lambda$4026, HeI $\lambda$4388 and HeI $\lambda$4922 being more easily detected in absorption.
Thus, a method to infer the presence of a starburst can be through the detection of the
higher order terms of the Balmer series and the HeI lines  in absorption. 

In fact, three (NGC 5135, NGC 7130 and IC 3639) of our four S2 that  show stellar wind
resonance lines in the UV, also show the higher order terms of the Balmer series and HeI 
($\lambda$3819, 4387 and 4922) in absorption (GD98). These results thus prove that the
high order lines of the Balmer series and HeI photospheric lines constitute a good indicator
of the presence of massive stars in S2 nuclei. In the other object, Mrk 477, that also
shows stellar wind resonance lines in the UV, the Balmer and HeI lines  are overwhelmed
by the corresponding nebular lines in emission. This implies that this method will not directly
detect the existence of massive stars in those nuclei for which the nebular lines are
strong and the starburst continuum is weak; thus, a lower limit to the real fraction of
objects that show starburst characteristics will  be inferred with this method. On the
other hand, Mrk~477 shows very broad emission at 4680 \AA\ due to Wolf-Rayet stars. 
Starbursts go through the Wolf-Rayet phase when stars more massive than 40 M$\odot$
evolve from the main sequence a few Myrs after the onset of the burst. 

Another important result from the analysis of the optical spectra of these four S2 nuclei
is that the metallic stellar lines due to old, cool stars are very diluted toward the nucleus. As
pointed out by CF98, a strong dilution of these lines happens only when the Seyfert
nucleus shows starburst characteristics.\footnote{A change in the metallicity of the stellar population
will also decrease the equivalent width of the metallic lines; however a strong dilution of these lines
is more probably associated to a change of the age of the nuclear stellar population than to metallicity,
because the metallicity will increase toward the nucleus when there is a gradient.} 

Thus, our method to infer the existence of starbursts in S2 galaxies is based mainly on
the detection of: 1) The higher order of the Balmer series in absorption and HeI lines;
2) Wolf-Rayet features at optical wavelengths;  3) strong dilution of the metallic lines
toward the nucleus.

\section{Results and interpretation}

\subsection{General considerations}

In this and the following two sections, we present the general results obtained from the
analysis of the spectra observed in the blue spectral range. The techniques used to
determine the stellar population in the nuclei of the S2  of our sample and the possible
contribution of the FC are these:

1) We compare the nuclear spectral energy distribution and the strength of the stellar lines
with  those in the spectra of the templates that we observed (NGC 205, NGC 1569, and NGC 4339)
and with off-nuclear spectra corresponding to the circumnuclear region of the galaxy
(section 5.2).

2) We trace the spatial distribution of the equivalent widths (Ew) of the stellar
metallic lines near the nucleus (section 5.3).
 
3) We model the photospheric HeI and Balmer lines with evolutionary synthesis models (section 5.4).

To perform this analysis we have extracted one-dimensional spectra from the
 two-dimensional frames.  For the extractions, we use the spatial distribution of the
continuum at 4550 \AA\ along the slit. In a few objects, the maximum of the continuum
is not coincident with the maximum of the distribution of nebular lines. The nuclear
spectrum corresponds to the central 3 pixels ($\sim$ 2 arcsec) around the maximum of the
continuum. The number of off-nuclear spectra varies from 2 in Mrk 78 and Mrk 463E
to over 30 in NGC 1068. These spectra cover a distance that ranges from $\pm$2 arcsec to
$\pm$60 arcsec. In these extractions, we add 3 or more pixels to get a signal to noise
(S/N) in the continuum better than 15. At the nucleus, the S/N is always better than 30.
To measure the strength of the stellar lines we have followed  the method of CF98 that
consists of determining a pseudo-continuum in selected  pivot-wavelengths and measuring
the equivalent widths (Ew) of the stellar absorption lines integrating  the flux under
the pseudo-continuum in defined wavelength windows. For the metallic lines, we use the
same as CF98 (for CaII K 3908-3952 \AA\ and G band 4284-4318 \AA) and for the H Balmer
and HeI lines those defined by  Gonz\'alez Delgado et al. (1999). We do not include in
our analysis CaII H and MgI+MgH because these lines can be strongly affected by nebular
emission (CaII H by H$\epsilon$ and [NeIII] and the MgI+MgH by FeII). The
pivot-wavelengths that we use are at 3780, 3810, 3910, 4015, 4140, 4260, 4430, 4800,
4905 and 4935 \AA.    

A glance at the nuclear spectra reveals that our targets show a variety of stellar
population properties, but according to their general continuum morphology they can be
grouped in four classes:  1) objects that have stellar absorption features that are strongly diluted at
the nucleus;  2) objects in which the stellar lines are similar to those of an old
stellar population;  3) objects in which the stellar lines are similar to a young and
intermediate age stellar population; 4) objects in which the stellar lines are similar
to an intermediate age stellar population. 

The detection and characterization of a young or intermediate age population will be
most difficult in those cases in which the spectrum is dominated by light from old stars,
and those cases in which the nebular emission-lines are very strong (and will therefore
fill in the relevant stellar absorption features). To summarize these constraints, and 
to illustrate the above classification scheme, in Figure 1
we plot the strength of a nebular emission-line (Ew(H$\beta$)) versus the strength
of the tracer of the old stars ((Ew(G band)).
The nuclei of groups 1 and 2 are located in different
places of the plot  
than nuclei belonging to the third and fourth groups. However,
we also found transition objects between groups.

\subsection{Empirical comparisons}

In this sub-section we report the results of the first technique described above. It allows us to derive the
general characteristics of the stellar population and to infer whether it is similar to
that of an elliptical galaxy or, in contrast, to younger stellar systems; and to
estimate the possible AGN contribution. We will discuss each of the four classes in turn.
       
\subsubsection{Nuclei with very weak stellar absorption-lines}

In this group are Mrk 477 and Mrk 463E. They both have very strong nebular H$\beta$ 
emission and the  stellar lines are very weak. The nucleus of Mrk 477 shows a broad emission
feature at 4680 \AA\ (Figure 9 in H97).  As has been discussed extensively by H97, the
origin of this feature could be due to Wolf-Rayet stars. We rule out the possibility
that this feature is scattered HeII $\lambda$4686 emission from the hidden  S1 nucleus
because the polarized spectrum does not show HeII emission, and the degree of polarization 
is only a few percent (Tran 1995b). Evidence that Mrk 477 harbors a nuclear starburst comes
from  the detection of the UV resonance wind stellar lines. This detection gives support to
the hypothesis that the  broad emission feature at 4680 \AA\ is produced by Wolf-Rayet stars.
Additional evidence comes from the  spectral energy distribution of the
nuclear continuum at optical wavelengths, which is well fitted by a starburst in the Wolf-Rayet
phase plus the contribution of the nebular Balmer recombination continuum (see Figure 6 in
H97). However, the Balmer and HeI photospheric absorption features in the nuclear spectrum
of Mrk 477  are completely overwhelmed  by the corresponding nebular lines. In contrast, a
circumnuclear spectrum of Mrk 477 extracted 2 arcsec from the peak of the optical
continuum  shows the higher order terms of the Balmer series in absorption. The spectrum is
also similar to that of the  companion galaxy of Mrk 477, that according to its nebular
line emission can be classified as a LINER (De Robertis 1987), but its continuum  shows a
starburst morphology (Figure 2). We will show later in Section 7 that the stellar
population of the  circumnuclear region of Mrk 477 is dominated by young and intermediate
mass stars. However, evidence for the  Balmer lines in absorption in the nuclear spectrum of
Mrk 477 is found indirectly. Using the profile of the nebular emission of H$\beta$ in the
nucleus, the Balmer decrement and the reddening that affects the ionized gas (E(B-V)=
0.23, H97), we have predicted the intensity of the nebular emission of H8-H10.  We have
subtracted these lines from the nuclear spectrum, and the resulting spectrum has H9 and H10 
in absorption. The Ew of these lines are $\sim$ 2 \AA, which is the typical value predicted
for very young  starbursts (Gonz\'alez Delgado et al. 1999). Thus, we conclude that the
circumnuclear region of Mrk~477 contains stellar features characteristic of a young and
intermediate age population.

As it has been pointed out before, the optical nuclear spectrum of Mrk 436E is very similar
to the nuclear spectrum of Mrk 477. Mrk 463E shows also a broad emission feature at 4680
\AA\ (Figure 3). To investigate the origin of this feature we have examined its
polarized optical spectrum. It shows broad H$\alpha$, H$\beta$ and FeII emission, but  it
does not show HeII $\lambda$4686 broad emission (Tran 1995b). Thus, as in Mrk 477, we can
rule out scattered light from the hidden S1 nucleus as the origin of the broad emission at
4680 \AA. However, this result by itself cannot prove whether or not there is a starburst in
the Wolf-Rayet phase in  the nucleus of Mrk 463E. To look for evidence of the evolutionary
state of the stellar population in Mrk 463E, we have predicted the intensity of H8-H10 in
emission as we did for the nuclear spectrum of Mrk 477.  In this case the higher
order Balmer series lines are only marginally detected, but with Ew's that are similar
to the case of Mrk 477.  The detection is not firm because the continuum in
the circumnuclear region of Mrk 463E is fainter than in Mrk 477 and the S/N is not good enough to
safely infer the stellar population. 
While the strong similarity
of the spectra of Mrk 477 and Mrk 463E is very suggestive of a starburst in the latter, it will have to be
confirmed at space-UV wavelengths.

In the absence of space-UV spectroscopy, images can alternatively
provide indirect evidence of a nuclear starburst in Mrk 463E through  the morphology of the
emission. We have observed Mrk 463E with HST and the Faint Object Camara (FOC) with the
F/96 relay that gives a spatial scale of 0.0014 arcsec/pixel. We used the F210M filter, because its bandpass
does not include any strong emission lines and so is dominated by the UV continuum (Figure 4). Mrk 463E
shows a UV knot which nature is unknown,  but most of its UV flux comes from 1 kpc extended
emission whose morphology is suggestive of the scattered light from the hidden  S1 nucleus known from
spectropolarimetry. Therefore,
we found that the UV morphology of Mrk 463E looks different from that of the other S2 nuclei
that directly show starburst characteristics in their UV spectra (see Figure 1 in H97 and Figures 4-7
in GD98). Thus, the origin of the continuum and the broad emission at 4680 \AA\  is
still an enigma. Mrk~463E is a very challenging object and HST+STIS observations will help to
further probe the origin of the broad emission feature at 4680 \AA\ and the FC.
 
\subsubsection{Nuclei with a dominant old stellar population}

In 8 of the objects (Mrk 3, Mrk 34, Mrk 348,
Mrk 573, NGC 1386, NGC 2110, NGC~5929 and NGC 7212) the stellar lines are similar to those of an
old population. The equivalent width (Ew) of the CaII K ($\lambda$3933) and the G band
($\lambda$4300) ($\geq$ 8 \AA\ and $\geq$ 4~\AA, respectively) are larger
than those in our young (1.8 \AA\ and 1.5 \AA\ in NGC 1569) and
intermediate age population templates (5.6 \AA\ and 3.9 \AA\ in NGC 205). However, the lines
are  weaker than in the elliptical galaxy (14 \AA\ and 7.5 \AA\ in NGC 4339). In particular,
in three of these S2 nuclei (Mrk3, NGC 7212 and Mrk 34) the stellar lines are significantly diluted
with respect to the elliptical galaxy. Figure 1 shows that these 8 nuclei are
distributed along a line in the plane Ew(H$\beta$)-Ew(CaII K), where the stellar lines
are weaker in those  objects with larger Ew of
H$\beta$ emission. This relationship suggests that the S2 nucleus that has a more powerful
engine source (thus, it has stronger nebular emission) can also have stronger contributions
of the scattered or reprocessed light from the hidden nucleus, and in consequence the stellar lines are more
diluted with respect to those in typical old stellar systems. Alternatively, as suggested by
S98, the stellar lines can be weaker than those in an elliptical galaxy because the stellar
bulge population in these objects has a larger fraction of intermediate age stars or the metallicity is
lower than in the elliptical galaxy. However, the metallicity is not the relevant parameter because it is
solar or over in the central regions of Seyferts (Gonz\'alez Delgado \& P\'erez 1996; Storchi-Bergmann
et al 1998), and thus it is similar to the metallicity in elliptical galaxies.

To distinguish between these two alternative solutions we have proceeded in two different
ways.  First, we fit the continuum distribution with our template elliptical galaxy plus 
the FC represented by a power-law (F$_\lambda\propto\lambda^{-1}$). Second, we fit the
nuclear continuum with an off-nuclear spectrum of the galaxy bulge as a template. The
goal is to check whether in this second way, the fit requires the presence of a FC as is
required using the elliptical galaxy as a template. In both ways, we add the contribution 
of the nebular Balmer continuum only if Ew(H$\beta$) is larger than 20 \AA\ because in weaker nuclei (Ew(H$\beta$)
$\leq$ 20 \AA) the nebular continuum contributes less than 2$\%$ to the total light at 4800 \AA. 

The results of the first exercise indicate that all the nuclei 
in this group are well fitted by the elliptical galaxy plus a power-law that  
contributes from 0$\%$ to 25$\%$ to the total light at 4800 \AA\ (the lowest contribution 
is in the nucleus of NGC 1386 and the largest one is in the nucleus of Mrk 34). 
The results of the second exercise indicate than   
in half of these objects (Mrk 348, NGC 1386, NGC 2210 and NGC 5929) the fit does not need the contribution of a
power law because the stellar lines are similar to those in their corresponding off-nuclear spectrum and the 
nuclear continuum distribution is redder than in the off-nuclear spectrum. Therefore, 
the small dilution of 
the nuclear stellar lines in Mrk 348, NGC 1386, NGC 2210 and NGC 5929 
with respect to the elliptical galaxy
is most probably related to a change of the stellar population with respect to that of 
NGC 4339. The cause can be a larger contribution 
of intermediate age stars than in an elliptical galaxy. 
In the other four nuclei (Mrk 3, Mrk 34, Mrk 573 and NGC 7212) the fit using the elliptical galaxy spectrum is better than
that using the off-nuclear spectrum. Thus, the dilution of the stellar lines in these nuclei with respect to 
the elliptical galaxy is probably related to the presence of scattered light produced by the hidden Seyfert~1 nucleus.
Examples of the fits are in Figure 5.

Therefore, we conclude from this analysis that the stellar population of the nuclei of
galaxies in this group is, in general, similar to that of an elliptical galaxy but they can
also have a larger contribution of intermediate age stars. Thus, the characterization of the
nuclear stellar population is important to infer the possible
existence of scattered light from the hidden Seyfert source and  its contribution to the
optical light. The comparison of the nuclear spectrum with the bulge spectrum as well as
with galaxy templates can help to provide a good estimate of the FC if it exists.

\subsubsection{Nuclei with young and intermediate age populations}

HeI and/or the higher order terms of the Balmer series have been directly detected in 6 nuclei of the
sample. They are: Mrk 273, Mrk 1066, Mrk 1073, NGC 5135, NGC 7130 and IC 3639 (see Figure
6). They have relatively weak nebular emission (Ew(H$\beta\leq$ 25 \AA)
allowing the direct detection of some photospheric Balmer and HeI lines. They show the G band and 
CaII K lines weaker than 4 \AA\ and 7 \AA, respectively. Thus,
the stellar population of these nuclei is very different from the stellar population in
elliptical  galaxies and it is more similar to young and intermediate age stellar
systems. A comparison of the nuclear spectra with the corresponding off-nuclear spectra indicate
that the nuclei of NGC~5135, NGC 7130 and Mrk 1066 are bluer; Mrk 273
and Mrk 1073 have similar colors to their off-nuclear spectra; and the nucleus of IC 3639 is redder
than its off-nuclear spectrum. In all the nuclei of this group, the stellar lines are
diluted  with respect to the off-nuclear spectra. This dilution could be due to the
presence of scattered light from the nucleus or by a larger contribution of young stars
in the nucleus.

We have tried to fit the spectral distribution of the continuum and stellar features combining the
off-nuclear bulge spectra and a power-law, as we did in Section 5.2.2. We found that we need
contributions of the FC at 4800 \AA\ of 50 $\%$, 40 $\%$ and 30 $\%$ for NGC 5135, NGC 7130 and
Mrk 1066, respectively (Figure 7). However, this result is not supported by our UV 
spectroscopic
observations of NGC 5135 and NGC 7130. If the FC contributed to the optical light with
these large fractions, then the UV continuum of these nuclei
would be totally dominated by scattered light from the hidden S1 nuclei, and the spectra
would not show any stellar features. In contrast, the UV spectra of NGC 5135 and NGC 7130
clearly show strong resonance lines formed in the wind of massive stars; the UV
wavelengths are totally dominated by a very powerful and young starburst (GD98). The reason
why the optical continuum can be fitted by the off-nuclear spectrum plus 
 the contribution of the power-law is
because young starbursts have spectral energy distributions that mimic the shape of a power-law
(Cid Fernandes \& Terlevich 1995). Therefore, these results point to an alternative
interpretation, namely, that young powerful starbursts contribute significantly to the
optical light of the nuclei of NGC 5135, NGC 7130 and Mrk 1066.

The optical continua of the nuclei of Mrk 273 and Mrk 1073
are well fitted by the off-nuclear spectra, but their stellar lines are weaker in the
nuclei. These results also suggest that young
starbursts also contribute to the optical nuclear light of Mrk 273 and Mrk 1073,
but they are very reddened. Therefore, these red starbursts contribute by diluting the
stellar lines but do not make the nuclear spectra very blue with respect to the off-nuclear spectra.

In this group, IC 3639 has the nucleus with the strongest G band and CaII K lines;
however the Ew of H$\beta$ in emission is similar to other objects in this group. The
strength of its metallic stellar lines are similar to those in the spectrum of an
intermediate age population system (e.g. NGC 205). Direct evidence that IC 3639 harbors 
a nuclear starburst comes from its UV spectrum. As in NGC 5135 and NGC
7130, the UV nuclear spectrum of IC 3639 shows stellar resonance lines formed in the winds of
massive stars (GD98). The starburst in the nucleus of IC 3639 is dusty and less luminous
than those in NGC 5135, NGC 7130 and Mrk 477. Therefore, we expect that its contribution to the optical light is
lower than in the latter objects. A starburst weaker than in NGC 5135 and NGC 7130 
will also explain why the metallic stellar lines in IC 3639 are less diluted than in NGC 5135 and NGC
7130 and why the spectrum is not so blue. We will see later that the Balmer absorption lines
are fitted with the contributions of a young and an intermediate age population. 

In Mrk 533 and Mrk 1 the stellar lines due to old, cool stars are similar in strength to the objects in the above
group, suggesting a similarly strong contribution from young/intermediate age stars. However, the nebular line emission
in these nuclei is substantially stronger, making
the direct detection of the stellar Balmer absorption lines more difficult. However, we do find 
evidence for 
the existence of young and intermediate age stars in the circumnuclear region of these galaxies,
as we now discuss in detail.

The metallic stellar features in Mrk 533 (EW(G band)= 1.9 \AA\ and EW(CaII K)= 3.8~\AA) are very 
diluted with respect to those of an elliptical galaxy, and its continuum is bluer than that 
of elliptical galaxies. We find that it is not possible to adequately fit the spectral energy 
distribution of the optical continuum with only a combination of an elliptical galaxy plus a power-law. 
The strength of the stellar lines is similar to the more diluted objects of the third group 
(e.g. of NGC 5135, NGC 7130 and Mrk 1066). This suggests that the stellar population and the spectral 
energy distribution in Mrk 533 could be better reproduced by a young or intermediate age stellar population. 
A comparison with the continuum distribution of NGC 205 shows that the nucleus of Mrk 533
is bluer than NGC 205; thus, a contribution of a power-law or/and a younger stellar population than that of NGC 205
is required to fit the shape of continuum in the nucleus of Mrk 533. In fact, a power-law that contributes
 15$\%$ of the total light at 4800 \AA\ plus the spectrum of NGC 205 can reproduce the
continuum, except below 3750 \AA\ where Mrk 533 is bluer than the composite spectrum.
However, this result does not prove the existence  of scattered AGN light in the nucleus of
Mrk 533, because a combination of two off-nuclear spectra  extracted $\sim$ 5 arcsec
from the peak of the continuum also fits the nucleus of Mrk 533 well if the off-nuclear
template is reddened by E(B-V)= 0.1. This template shows Balmer absorption lines, indicating
that an intermediate age population contributes significantly to its optical light. However,
the nucleus of Mrk 533 has stronger nebular Balmer lines than the other nuclei of the third
group, and the direct detection  of the photospheric lines is more difficult than in the
cases of Mrk 273, Mrk 1066, Mrk 1073, NGC 5135, NGC 7130 and IC 3639. 

To detect indirectly the Balmer lines in absorption, we have predicted the intensity of the 
H$\delta$-H10 by scaling
the profile of H$\beta$ using the Balmer decrement and the extinction, E(B-V)= 0.3, derived from the ratio 
H$\gamma$/H$\beta$. We subtract the predicted nebular lines from the nuclei of Mrk 533 and the resulting spectrum shows
absorption features that are very similar to those in the off-nuclear template (Figure 8). 

The nucleus of Mrk 1 shows metallic stellar features that are very diluted with respect to those in an elliptical 
galaxy and are similar to objects in the third group. Its continuum distribution 
cannot be fit by the combination 
of a power-law and an elliptical galaxy because the
power-law has to  contribute 40$\%$ of the total light at 4800 \AA\ to reproduce the
shape of the continuum, and the stellar lines in the composite fit spectrum are then weaker 
than in the nucleus. However, the nuclear spectrum can be fitted by an off-nuclear spectrum
if it is reddened by E(B-V)= 0.2. The spectrum is extracted  from 3 arcsec off the continuum
peak at 4450 \AA. The shape of the continuum and the stellar features in
the circumnuclear spectrum are similar to a spectrum dominated by an intermediate age
population, with the high order Balmer series in absorption. However, these lines
are not detected in absorption in the nuclear spectrum because the nebular contribution is
very strong. Mrk 1 has EW(H$\beta)\sim$ 70 \AA, and is one of the nuclei with the strongest nebular
Balmer emission (after Mrk 477 and Mrk 463 E).

To indirectly detect these absorption lines, we predict the nebular H$\delta$-H10 in the same way as we did in
Mrk 477, Mrk 463E and Mrk 533. We subtract the nebular lines from the nuclear spectrum of Mrk 1.
The resulting spectrum
has absorption features that are very similar to those in the off-nuclear spectrum (Figure 9). 

We also find (as in Mrk 477 and Mrk 463E) that there is a weak, very broad emission feature underlying the HeII$\lambda$4686
nebular emission line in Mrk 1 (Figure 10). As in the other cases, we cannot prove that this feature is due
to a substantial population of Wolf-Rayet stars. We can however rule out an origin due to scattered AGN
light, since this broad bump is not seen in the polarized-flux spectrum. 

To conclude, our analysis strongly suggests that the nuclear stellar population of Mrk~533 and Mrk 1 has an important contribution 
of young and intermediate~mass stars, even though the direct spectroscopic signature of these stars is not clear. 

\subsubsection{Nuclei with a pronounced intermediate age population}

The metallic lines in the nucleus of Mrk 78 are diluted with respect to those in an
elliptical galaxy and are similar to those in NGC 205.
The nucleus also shows relatively weak nebular Balmer lines,  facilitating the detection
of the Balmer absorption lines. In fact, H$\delta$ and H$\epsilon$ show a hint of
absorption. Below 3900 \AA, the spectrum shows Balmer absorption features, but they do not
correspond to those of a young (a few Myr old) age population. Thus, the stellar
population in Mrk 78 consists of a mixture of old and intermediate age stars.
Indeed, a combination of a 1 Gyr old (A and F star dominated)
population with that of an  elliptical reproduces well the continuum shape and the
strength of the stellar lines (Figure 11).    

The nuclear spectrum of NGC 1068 corresponds to the central 2 arcsec around the continuum peak. This
spectrum shows that the metallic stellar lines are very diluted with respect to an old stellar
population. A combination of an elliptical galaxy and a  power-law can reproduce the
continuum shape if the power-law contributes with at least 35$\%$ of the total
light  at 4800 \AA, but this combination cannot fit the observed stellar lines. A
comparison of the nuclear with an off-nuclear spectrum, extracted 4 arcsec from
the continuum peak, indicates that the nucleus has also weaker metallic lines
than the off-nuclear spectrum. Thus, the contribution of the scattered light from the
hidden S1 nucleus can contribute to the dilution of these the lines. A combination of this off-nuclear
spectrum plus a power-law that contributes 10-15$\%$ to the total light at 4800
\AA\ fits it well (Figure 12). An inspection of this off-nuclear spectrum shows that the
stellar population there has an important contribution of intermediate age stars (see
section 5.3).

\subsection{Spatial variation of the stellar lines}

The spatial distribution of the Ew of the metallic lines around the nucleus allows us to
estimate the possible variations of the stellar population as a function of distance from the nucleus.
That is, the presence of a younger stellar population will produce a strong dilution of the
stellar lines toward the nucleus. However, dilution of the stellar lines can also be caused
by scattered light from the hidden S1 nucleus.  But the dilution factor, together with the
stellar characteristics of the nuclear and off-nuclear spectra, can point to one of
these two alternatives.

\subsubsection{Nuclei with very weak stellar absorption-lines}

In Mrk 477 the spatial distribution of the G band and the CaII K lines shows a deep dip in
the Ew of the lines at the nucleus (Figure 13). Comparing to the values out of the
nucleus, the dilution 
of the light of old stars is $\sim70\%$. 
This strong dilution of the stellar lines at the location of the continuum peak cannot
be produced by a power-law, because if that were the case the UV light would have to be
totally dominated by FC. On the contrary, we found that the UV light is produced by a
starburst in the nucleus of  Mrk 477 (H97).

In Mrk 463E we have extracted two off-nuclear spectra at $\pm$2 arcsec from the nucleus.
The Ew of the stellar lines in these off-nuclear spectra are 2.7 \AA\ and 1.5 \AA\ for the
CaII K and the G band, compared to 0.9 \AA\ and 0 \AA, respectively, on the nucleus.
Thus the metallic lines are significantly more diluted in the nucleus.

\subsubsection{Nuclei with a dominant old stellar population}
  
The plot of the EW of the metallic lines as a function of the distance from the nucleus
shows that the  variation of the strength of the stellar lines in the nucleus and
circumnuclear region of these galaxies is  small or null (Figure 14). The dilution factor is
$\leq 10\%$. The exception to this bevavior is Mrk 3, where the stellar lines at the
nucleus show a  significant dilution with respect to the circumnuclear region (Figure 14). 
The estimated dilution factor is $\sim 25\%$. 
 
The lack of any significant dilution in Mrk 573, Mrk 34 and NGC 7212 is surprising,
given that the continuum spectral energy distribution in these nuclei
required the contribution of a power-law.  As pointed out by S98, the presence of
an extended scattered continuum could be the explanation, because it can contribute equally
to the nuclear and to the off-nuclear spectra, thus keeping the EW of the lines
approximately at the same values.       

\subsubsection{Nuclei with young and intermediate age populations}

In the galaxies of this group, the Ew of the metallic stellar lines shows a spatial
variation, that indicates that there is a significant change of the stellar population with
distance from the nucleus. The metallic lines are diluted with  respect to the off-nuclear spectra; in particular,
the dilution is very strong in those nuclei that have bluer colors than their surroundings (Mrk
1066, NGC 5135 and NGC 7130). The dilution factor ranges between 45$\%$ and $\sim10\%$,
the stronger dilution corresponding to NGC 5135 and the lower to IC 3639.  The UV spectra of
IC 3639 and NGC 5135 both show strong evidence for the existence of a nuclear starburst
in these two galaxies. Thus, the significant difference between the dilution factor in NGC
5135 and IC 3639 could be related to the existence of a more powerful starburst in NGC 5135
than in IC 3639. Examples of the spatial variation of Ew  are in Figure 15.

\subsubsection{Nuclei with a pronounced intermediate age population}

In Mrk 78 we have extracted two off-nuclear spectra that are at $\pm$2 arcsec from the
nucleus. Even we have only three points, the plot of Ew of the stellar lines does not show
a variation of the strength of the stellar lines with the distance from the nucleus.

In contrast, in NGC 1068 we have traced the spatial distribution of the strength of the
stellar lines out to $\pm$60 arcsec from the nucleus. There is a significant change of the Ew
of the lines with the distance.  The nuclear population is strongly diluted with respect to the
population at $\pm$5 arcsec. Values of the Ew of the  lines at $\pm$10 arcsec are
similar to those of the nucleus; however, the dip of the strength of the lines is due to the
presence of young massive stars in the circumnuclear ring (Figure 16).  

\subsection{Modeling the Balmer absorption lines}

In the previous analysis we have found that  six nuclei (Mrk 273, Mrk 1066, Mrk 1073, NGC
5135, NGC 7130, and IC 3639) have strong Balmer absorption lines. These lines are also
detected in the circumnuclear regions in another three galaxies (Mrk 533, Mrk 1, and Mrk
477) and in their nuclei when the contribution of the nebular emission is subtracted. The
strength of these and the metallic lines indicates that the stellar population is dominated
by young and intermediate mass stars. We also found that the circumnuclear region of NGC
1068 and the nucleus of Mrk 78 show Balmer lines, but 
the strength of the Balmer and metallic lines suggests that the stellar
population in these two objects has an important contribution of intermediate age stars.

To estimate the contribution of young and intermediate~mass stars to the optical light of
these objects, we use the technique of stellar evolutionary synthesis. The models predict
the profile of the HeI and Balmer absorption lines for single metallicity stellar population
as a function of the age, assuming an initial mass  function (see Gonz\'alez Delgado et al.
1999 for details). These models  use a stellar library that comprises synthetic spectra of
stars that were generated with stellar atmosphere models  (see Gonz\'alez Delgado \&
Leitherer 1999 for details). We have updated these models to predict several spectral ranges
that were not computed previously. Thus, these new models predict the CaII H and K,
H$\epsilon$ and the spectral range around the G band. The stellar spectra reproduce well
the stellar lines with effective temperature larger than 6000 K; thus, they can be used to
predict the photospheric stellar lines of populations younger than a few Gyr old. 

To avoid having to introduce the extinction as an additional free parameter in the fit to the S2 spectra, we
use normalized spectra, by dividing by the continuum resulting of fitting a line between two
consecutive pivot wavelengths (defined in section 5). A general result is
that the stellar population cannot be described by only two components, for example an
old (10-15 Gyr) plus a young or intermediate population. Thus, the fit of the metallic
and Balmer lines requires three components: a young, an intermediate and an old population.
We take for the young population an instantaneous burst 3 Myr old, for the intermediate age
population a combination of three bursts of 200, 500 and 1000 Myr old (roughly equivalent 
to a burst lasting 800 Myr commencing 1 Gyr ago), and for the old population the elliptical template. 
To show the validity of the process followed, we present first the result of fitting  our intermediate age
stellar population galaxy template NGC 205. We find a good fit with a combination of  
intermediate age (0.2 to 1 Gyr) plus the elliptical galaxy if they contribute to the light at
4800 \AA\ with 70$\%$ and 30$\%$, respectively (Figure~17). This solution reproduces quite
well not only the strength of the lines but also the shape of the continuum. It agrees with
that found by Bica et al. (1990). We report the results of the fit by grouping the
objects according to the contributions of the young and old populations. First, we report
the results of fitting the nuclear spectra (Figure 18), and then the off-nuclear spectra of
Mrk 1, Mrk 533, NGC 1068 and Mrk 477. The plots are in Figure 19.

\subsubsection{The nuclei of NGC 5135 and NGC 7130}

These spectra 
show very weak metallic lines and strong Balmer and HeI ($\lambda$3819 and $\lambda$4026) in absorption.
They both require an important contribution of a young stellar population. The strong
dilution of the G band and the CaII K line indicates that the old population contributes
very little to the optical light ($\sim 10\%$). The dilution of these lines  is produced by
the contribution of a young and an intermediate-age stellar population that provide
50$\%$ and 40$\%$, respectively, of the total optical light (see Figure 18a).

\subsubsection{The nuclei of Mrk 273, Mrk 1066 and Mrk 1073}

The strength of the metallic lines requires a contribution of the old population that is 
larger than in NGC 5135 and NGC 7130; 
it is between 20$\%$ (for Mrk 1066) and 30$\%$ (for Mrk 273 and Mrk 1073). The strength 
of the Balmer lines requires an 
intermediate population that contributes 50$\%$, and a young burst that contributes
30$\%$ (in Mrk 1066) and 20$\%$ (in Mrk 273 and Mrk 1073) (see Figure 18b and c).
However, we have noticed that a similar fit is obtained if we substitute the contribution of
the young stellar population by the contribution of a power-law that provides 
30$\%$ and 20$\%$ of the total optical light.

\subsubsection{The nucleus of IC 3639}

IC 3639 has the nucleus with the strongest metallic lines of its group, and the fit of the lines requires a 
contribution of the old stellar population that is the largest fraction, 40$\%$ of the total optical light. 
The intermediate and the young stellar
population contribute 40$\%$ and 20$\%$ of the optical light, respectively (Figure 18d). 

\subsubsection{The nucleus of Mrk 78}

As we already explained above, the nucleus of Mrk 78 shows Balmer absorption lines, but
they are weaker than those expected for an intermediate age population. The strength of the G
band and the CaII K lines  suggests that the contribution of the old population is important. The
combination of the intermediate age population with  the elliptical galaxy fits the nuclear spectrum
reasonably well
if each population contributes half of the total optical  light (Figure 10).
However, from the fit of the stellar lines, we cannot rule out that 10$\%$ of the total
light could be provided by a power-law. 

\subsubsection{Off-nuclear spectra of Mrk 1 and Mrk 533}

The nuclei of these two objects have the stellar Balmer lines filled in by the corresponding
nebular lines.  However, as we have already seen, the nuclear spectra show the absorption
features when the nebular contribution is subtracted. The resulting spectra have
features similar to the circumnuclear spectra with characteristics of an
intermediate age population. To perform the fit we use the off-nuclear spectra of Mrk 1 
extracted 3~arcsec from the continuum peak and one of Mrk 533 at 5 arcsec. Both spectra
are well fitted  by a combination of young, intermediate and old populations that contribute
20$\%$, 40$\%$ and 40$\%$, respectively (Figure 19a).

\subsubsection{Off-nuclear spectra of NGC 1068}

We have fitted two circumnuclear spectra of NGC 1068. One is the average of two spectra (extended
2 arcsec) extracted  at $\pm$ 4 arcsec of the continuum peak at 4450 \AA, that represents
the off-nuclear stellar population of NGC 1068.  We call it the bulge spectrum. The second spectrum
corresponds to the circumnuclear HII ring and it is extracted from  8~arcsec of the
continuum peak and it extends 7 arcsec. We call it the HII ring spectrum. 

The bulge spectrum is well fitted by combining an intermediate age population with the elliptical galaxy,
each contributing half to the total optical light
(as in Mrk 78). A power-law that contributes 10$\%$ and the rest provided by the intermediate age
plus the old population is 
also acceptable (Figure 19b). 
The HII ring spectrum requires the contribution of three populations, a young burst, an intermediate and an old 
stellar population. 
They contribute  40$\%$, 30$\%$ and 30$\%$ of the total optical light, respectively (Figure 19c). 

\subsubsection{Off-nuclear spectrum of Mrk 477}

We take an off-nuclear spectrum 1.5 arcsec from the continuum peak and extended by 2
arcsec.  The spectrum clearly shows the higher order terms of the Balmer series in
absorption, and the G band and  Ca II lines. We fit also the spectrum with the contribution of
three components, a young burst, an intermediate  and an old stellar population, 
that contribute 30$\%$, 40$\%$ and 30$\%$ of the total optical light respectively, (Figure 19d).

\subsection{CaII triplet: general results}

The most prominent stellar feature in the red/near-IR region is the CaII triplet ($\lambda$8498, 8542, 8662).
These lines are  detected in absorption in many Seyfert galaxies (Terlevich et al. 1990; Nelson
\& Whittle 1995), in most of  the nuclear starburst galaxies (Storchi-Bergmann et al, 1995) and
in the spectra of young stellar clusters (Bica, Santos \& Alloin 1990; Gonz\'alez Delgado et al
1997). These lines are formed in the photospheres of asymptotic giant branch,
red giant  and red-supergiant stars 
(D\'{\i}az,
 Terlevich \& Terlevich 1989). They are very strong in starbursts with ages between 10 and 20
Myr (Mayya 1997; Garc\'\i a-Vargas, Bressan \& Moll\'a 1998) due to the evolution of massive 
stars to the red supergiant phase.

We have observed our sample in the red to detect the CaII triplet and to estimate the possible
dilution  of these lines with respect to normal galaxies. We also observed our template
galaxies. Three of our targets  (Mrk 533, NGC 5135 and IC 3639) could not be observed. The two most
distant objects of our sample (Mrk 463E and Mrk 34) were affected too strongly by the strong sky
emission at 9000 \AA\ and longer wavelengths. However, in the remaining objects the lines are
very strong. We have measured the equivalent widths of the lines using the line windows defined
by D\'\i az et al (1989); the values are in Table 3. 

We found that the nuclei that belong to the second group (i.e. those with a stellar
continuum dominated by an old stellar population) show a CaII triplet strength that is very similar to 
the elliptical galaxy NGC 4339 (Figure 20a). The exceptions to this behavior are Mrk 3 and NGC
7212. They both have CaII triplet equivalent widths lower than the
elliptical galaxy and than the rest of the nuclei of this group. These two nuclei have very
strong nebular emission and the blue stellar lines are diluted with respect to an
old population. We found that the dilution in the blue spectral range is due to the contribution
of scattered light from the hidden Seyfert
nuclei plus the nebular continuum. Thus, these two contributions can also dilute the CaII
triplet. On the other hand, these two nuclei have very strong Balmer nebular emission. Thus, 
Pa13 and Pa16 that fall at the same wavelengths as CaII $\lambda$8452 and 
CaII $\lambda$8662 can partially fill in the absorption lines. Mrk 3 and NGC 7212 also show
a broad emission feature at 8600 \AA\ that is produced by
emission of [ClII] $\lambda$8580, Pa 14, and [FeII] $\lambda$8617.

The nuclei that show the higher order Balmer series in absorption also show strong CaII
triplet lines that have equivalent widths similar to those of NGC 205 (Figure 20b). The nucleus
of Mrk 1 also has the CaII triplet with strength similar to that of NGC 205. However, these
lines in NGC 205 are only slightly weaker than those in the NGC 4339. The CaII triplet has been
detected also in the nuclear spectrum of Mrk 477 (one of the two objects with a blue spectrum
that does not show any stellar features) with a strength similar to that in NGC 205. 

\section{Discussion}

\subsection{Properties of the sample in the infrared}

The infrared emission from S2 galaxies comes three different sources. Dust in/near the central
engine (probably associated with the dusty obscuring torus) will be heated by a hidden S1 nucleus
and emit primarily in the near- and mid-IR. Dust heated by a hot starburst population will emit primarily
in the far-IR, as will the dust heated by the normal stellar radiation field in the large-scale
ISM. In this section, we will assess the origin of the infrared emission in our sample and
compare the infrared properties of the S2 nuclei with predominantly old and young/intermediate
age stellar populations. The infrared dominates the broad-band spectral energy distributions
of nearly all our galaxies.
The total infrared luminosity estimated with the fluxes of all four IRAS bands
(Sanders \& Mirabel 1997) ranges from 3$\times10^{9}$ to 1.2$\times10^{12}$ L$\odot$ for our sample.
All the galaxies, except NGC 1386, have an infrared luminosity larger than
the typical value for a normal spiral (10$^{10}$ L$\odot$), and the log-mean luminosity of
our sample is 10$^{11}$ L$\odot$. Therefore,
most galaxies in our S2 sample can also be classified as Luminous Infrared Galaxies.

The objects in our sample were selected based on the emission-line and/or radio 
continuum fluxes of their nuclei. They are therefore amongst the brightest S2 nuclei of the 
Whittle (1992) compilation. This is also true in the mid-IR. In particular,
all the galaxies in the sample are also
powerful infrared emitters at 12 $\mu$m, which Rush, Malkan, \& Spinoglio (1993) argue is the
least-biased waveband for selecting Seyfert nuclei.
The 12 $\mu$m luminosity for our sample ($\nu$$P_{\nu}$), ranges
from 6$\times10^8$ to 2$\times10^{11}$ L$\odot$. Most of the galaxies have L$_{12}\geq10^{10}$
L$\odot$, and only three galaxies (NGC 1386, NGC 2110, and NGC~5929) have L$_{12}$ significantly
lower than 10$^{10}$ L$\odot$.
Bonatto \& Pastoriza
(1997) have compiled the IRAS fluxes of a large sample of Seyfert galaxies, and they
found that, for the S2, the L$_{12}$ distribution peaks at 10$^{10}$ L$\odot$. Thus,
the nuclei selected in our sample are amongst the most powerful and brightest S2 at mid-infrared
wavelengths. The S2 in our sample are much more luminous in the mid-IR than normal galaxies.
According to the luminosity function presented by Rush,
Malkan \& Spinoglio (1993), only a small fraction of normal galaxies have L$_{12}\geq10^{10}$ L$\odot$.
We found for our sample that L$_{12}$ correlates with the power of the AGN as measured by both
the [OIII] $\lambda$5007 luminosity and the 
1.4 GHz non-thermal nuclear radio continuum power. These results suggest that (contrary
to normal galaxies, where the 12 $\mu$m flux is emitted mostly by small grains in the large-scale
interstellar medium)
the 12 $\mu$m continuum in our S2 sample is emitted by warm dust heated by the AGN.

These results imply that our sample is representative of the brightest and most powerful S2 nuclei
in the local universe (regardless of how the nuclear luminosity is measured). In particular,
even though many of the galaxies in our
sample were originally discovered by their ultraviolet excess (e.g. in the Markarian 
surveys), most of them are also abnormally strong mid-infrared and radio emitters. Therefore, we can conclude 
that our sample is not biased towards nuclei that have a starburst, as 
could be thought {\it a priori} for objects initially discovered by their ultraviolet emission.  

We can compare the nuclear luminosities of 
the S2 nuclei with predominantly old and young/intermediate age stellar populations.
The 12 $\mu$m luminosity of the nuclei whose optical continuum is dominated 
by an old stellar population (Mrk 3, Mrk 34, Mrk 348, Mrk~573, NGC 1386, NGC 2110, 
NGC 5929, and NGC 7212) has a mean value of log L$_{12}$/L$_{\odot}$ = 9.84$\pm$0.19. In contrast,
those nuclei that have a young and intermediate age population (Mrk 1,
Mrk 273, Mrk 477, Mrk 533, Mrk 1066, Mrk 1073, NGC 5135, NGC 7130, and IC~3639) have a
mean log L$_{12}$/L$_{\odot}$ = 10.37$\pm$0.11. Thus, we find that the S2 in our sample
that show starburst characteristics may be more powerful AGN (at the 2.5$\sigma$ significance level). 
To test this
further, we have compared the luminosities of the [OIII] emission-line and the 1.4 GHz nuclear nonthermal
radio emission in the two sub-groups. The S2 group with a dominant old population has a mean
value log $L_{[OIII]}$/erg s$^{-1}$ = 41.51$\pm$0.32  compared to 41.55$\pm$0.21
for the S2 nuclei with young/intermediate
age populations. The S2 group with a dominant old population has a mean
value log $L_{1.4}$/erg s$^{-1}$ = 38.63$\pm$0.34,
compared to 39.09$\pm$0.16 for the S2 nuclei with young/intermediate populations. In all three
comparisons, the younger stellar populations are associated with more luminous AGN, but
the statistical significance ranges from nearly zero to marginal.
Larger samples, especially those that better sample the highest luminosity S2 nuclei,
will be required to confirm this.

A stronger difference between the old and young/intermediate S2 nuclei is found if we compare 
the far-infrared (40 to 120 $\mu$m) luminosity. The mean value in the first group of nuclei (old stars) is
log $L_{FIR}$/L$_{\odot}$ = 10.05$\pm$0.20. This is
significantly lower (3$\sigma$) than the emission in galaxies that have nuclei with young 
and intermediate age stars (log $L_{FIR}$/L$_{\odot}$ = 10.83$\pm$0.17). This result is very
reasonable since the FIR is an excellent tracer of dusty star-formation in galaxies. In the local
universe, galaxies with FIR luminosities greater than log $L_{FIR}$ $\sim$ 10.5 are almost exclusively
starbursts. We find that $\sim$80$\%$ of the young/intermediate S2 nuclei in our sample exceed
that threshold, compared to only 25$\%$ of the old S2 nuclei. These results suggest that there
is a direct connection between the young/intermediate age stars and the population that
heats the dust radiating in the far-IR. This has been confirmed quantitatively for
the four S2 nuclei studied in detail by GD98.

\subsection{Implications for the starburst-Seyfert connection}

The detection of the higher order terms of the Balmer series in absorption in the 
near-ultraviolet spectra in about half of the S2 of the sample, contrasts with the 
general view that the nuclear spectra of Seyferts are composed of an old stellar 
population plus the scattered light from the hidden nucleus. The analysis of the 
near-ultraviolet spectra indicates that young and intermediate age stars are present 
in the nuclear and circumnuclear regions of 45\%\ of the S2 galaxies of this 
sample\footnote{Storchi-Bergmann et al. (2000) have recently presented the analysis 
of a sample of 20 Seyfert 2 galaxies, of which 5 are in common with our sample, to 
look for signatures of young and intermediate massive stars, and they find a 
frequency of nuclear starburst similar to us.}  . 
In an additional 10\%\ the sample, an intermediate age population contributes significantly 
to the optical light. However, the contributions of young and intermediate age stars 
change from object to object. We find the largest contributions in the nuclei of 
NGC 5135 and NGC 7130, both with a fraction of 50\%\ of young and 40\%\ intermediate 
age stars. The lowest contributions are found in IC 3639, with fractions 
of 20\%\ and 40\%\ of young and intermediate age stars, respectively. Because massive 
stars are very blue and emit most of their flux at ultraviolet and far-ultraviolet 
wavelengths, the corresponding fractions of young stars will increase to shorter 
wavelengths. Thus, the ultraviolet spectrum of these nuclei will be dominated by the 
massive stars, even in the nuclei with a relatively low contribution of young stars
to the optical light (as
our HST-UV observations of IC 3639 show - GD98).

The contribution of young and intermediate age stars at optical wavelengths in these
nuclei are also very similar to the situation in
normal starburst nuclei. A prototypical nuclear starburst galaxy is 
NGC 7714 (Weedman et al. 1981). Its ultraviolet spectrum and nebular optical emission 
lines are very well explained by a 4-5 Myr old burst (Gonz\'alez Delgado et al. 1999).
In contrast, its optical continuum shows an important contribution of intermediate age
stars. A fit of its ultraviolet to near-infrared spectral energy distribution shows that
half of the optical light in the central 2$\times$2 arcsec is provided by a $\sim$200 Myr burst 
(Goldader et al. 2000). The near-ultraviolet spectrum of NGC 7714, just like half of the
Seyfert 2 nuclei in our sample, shows the higher order terms of the Balmer series in
absorption (Gonz\'alez Delgado 2000). The strength of these lines are well fitted by two
bursts, of 5 and 200 Myr, that contribute to the total light at 4400 \AA\ with a similar
fraction as in the cases of NGC 5135 and NGC 7130. The older burst in NGC 7714 could be
a consequence of the interaction with its companion, NGC 7715, because the dynamical time
scale of the interaction is $\sim$ few$\times$100 Myr (Gonz\'alez Delgado et al. 1995). 

From our comparison with NGC 7714, we conclude that half of the S2 of our sample show spectral
characteristics that are very similar to prototypical starburst nuclei. On the other hand,
the fact that 6 of the 9 S2 nuclei of the sample that show starburst characteristics are
interacting or merging systems, suggests that the $\sim$ few$\times$100 Myr burst could be
triggered by the close passage of a companion, such as the merger models of Mihos \&
Hernquist (1996) predict. According to these and other models (e.g. Athanasoula 1992) 
based on the existence of a non-axisymmetric potential, the gas sinks toward the 
central part of the galaxy. There, it could be compressed producing a starburst and 
feeding the active nucleus. 

Young and intermediate age stars have been detected in the optical continuum of other
active galaxies. Schmitt et al. (1999) find that the optical light of 30\%\ of the
nuclei in a sample of 20 S2 galaxies is dominated by young and intermediate age stars, and
80\%\ of the nuclei have also a significant contribution of 1 Gyr old stars. These stars
contribute to the optical light with larger fractions that they  do in normal early type
galaxies. Intermediate age stars have been found also in the nearby radio galaxy 3C 321
(Tadhunter, Dickinson \& Shaw 1996), and in high redshift radio galaxies selected
to be very strong far-infrared and radio emitters (Tran et al 1999). The most spectacular
case is found in the QSO UNJ1025-0040 (Brotherton et al 1999). Its ultraviolet spectrum
shows broad MgII $\lambda$2800, while in the near-ultraviolet a strong Balmer
absorption series is detected. This QSO could be interacting and the companion galaxy
shows also strong Balmer absorption series (Canalizo et al 2000).

Our results and those reported by other groups are pointing to a causal link between
the starburst and the Seyfert phenomenon. In this case, the few$\times$100 Myr star
formation time scale should be comparable to the duration of the nuclear activity. As
Storchi-Bergmann et al (1998) have pointed out, this timescale - $\sim1$\%\ of the Hubble
time - would imply that $\sim1$\%\ of the galaxies should be strongly active, as current estimates
indicate (e.g. Huchra \& Burg 1992). 

\subsection{Are there two kinds of Seyfert 2's?}

The results of the previous sections show that the nuclear and circumnuclear regions of
S2  have a variety of stellar population properties. However, most of the nuclei of our
sample can be grouped in two classes, those with young and intermediate age stars and
those with an optical continuum dominated by old stars. This segregation in two groups
poses an important question: are there two different classes of S2 nuclei? That is, do all S2
nuclei contain a hidden S1 nucleus? Do all S2 nuclei contain nuclear starbursts, but
undetectably faint in half the cases due to low relative luminosity?
We note in this vein that the significantly higher far-IR luminosities of
S2 nuclei with young and intermediate age stars (see above) probably rules out a model
in which the old-star-dominated S2 nuclei have equally powerful (but more heavily extincted)
starbursts.

We can begin to address the above questions by quantifying the relative energetic importance of
young stars and the hidden S1 nucleus in our sample. This cannot be done rigorously,
but there are some useful empirical diagnostics. The ratio of the mid- and far-IR luminosity
should probe the ratio of the S1 and starburst luminosities. Dust-reprocessed radiation
from the S1 nucleus peaks in brightness in the mid-IR (cf. Elvis et al. 1994), while starbursts
peak in brightness in the far-IR. Optical emission-line diagnostics are also potentially useful.
The characteristic excitation state of gas photoionized by the hidden S1 nucleus is very high
(e.g. [OIII]$\lambda$5007/H$\beta$ $\sim$ 10), while starbursts in the nuclei of spiral galaxies
have much lower excitation ([OIII]$\lambda$5007/H$\beta$ $\leq$ 1). Thus, the 
[OIII]$\lambda$5007/H$\beta$ ratio should contain information about the sources of ionizing
photons in the S2 nuclei.

Inspection of the infrared colors and excitation indicates that the two S2 groups do indeed have
different properties. The excitation measured by [OIII]/H$\beta$ is almost a
factor two lower in nuclei with young and intermediate age stars (mean value log([OIII]/H$\beta$
= 0.78$\pm$0.07)
than in nuclei with an old stellar population (mean value of 0.97$\pm$0.06). A similar result
is found for the infrared color f$_{25\mu}$/f$_{60\mu}$. Nuclei with young and intermediate
age stars are cooler (mean value of log(f$_{25\mu}$/f$_{60\mu}$) = -0.70$\pm$0.07) than nuclei with
an old stellar population 
(mean value of -0.40$\pm$0.09). These two quantities seem to be related: objects with higher
excitation are those with warmer colors (Figure 21). This is as expected if hot (young) stars
make a significant contribution to both heating dust and ionizing gas in some (but not all) S2 nuclei
(namely the S2 nuclei in which our spectra reveal a substantial young/intermediate age population).

An alternative interpretation of the mid/far-IR ratio in S2 nuclei is suggested by the results
of Heisler, Lumsden \& Bailey (1997). They found that the probability of detecting scattered broad
emission lines from a hidden S1 nucleus is highest in S2 nuclei that have warm infrared colors 
(f$_{25}$/f$_{60}>$ 0.25) and smaller extinction. According to the unified
scheme, S2 nuclei are those in which the molecular torus that surrounds the central
source is oriented with its equatorial plane near our line-of-sight. However, there should be S2
nuclei in which  the torus is oriented at intermediate angles between the equatorial plane and
polar axis. In these nuclei the S1's hidden Broad Line Region (HBLR) will be easily detected in
polarized light. At this
orientation we are seeing the inner part of the torus that is illuminated directly by the
central source. Because the temperature of the dust will be higher in the inner than in
the outer parts of the torus, the nuclei with intermediate orientation will be warmer at
infrared wavelengths than nuclei oriented with the polar axis perpendicular to us.
These results, together with the segregation of the nuclei of our sample in terms of
stellar population, excitation, and far-infrared colors, suggests a possible
connection between the detection of a HBLRs and the starburst activity in the nuclei of S2.

Most (18 of 20) of the S2 nuclei of our sample have been observed
spectropolarimetrically at optical wavelengths (Kay 1994; Tran 1995a; Heisler et al 1997).
They show an average optical continuum polarization of a few percent (mean $\sim2$\%),
which is much lower than the degree of polarization expected by the unified scheme
unless the continuum is primarily of some other origin (e.g. starlight). The
exception is NGC 1068, in which the continuum is polarized by $\sim16$\%\ (Antonucci \&
Miller 1985). Balmer broad emission lines have been detected in 8 of the nuclei in our
sample (Mrk 3, Mrk 348, Mrk 463E, Mrk 477, Mrk 533, NGC 1068, NGC 7212, and IC 3639, Tran
1995abc; Heisler et al 1997), and broad H$\beta$ is also marginally detected in Mrk 573 (Kay
1994). Heisler et al (1997) have also reported the non-detection of HBLRs in two of our
nuclei, NGC 5135 and NGC 7130. For the remaining objects, we have not found any
information in this respect in the literature. An inspection of Figure 21 indicates
that all S2 nuclei of our sample with HBLRs have IR colors warmer than 0.25. 

It is important to emphasize that HBLRs are present in at least 3 of the 9 S2 nuclei
in our sample with a young/intermediate age stellar population (Mrk 477, Mrk 533, and IC 3639).
This compares to 3 of the 8 S2 nuclei dominated by old stars (Mrk 3, Mrk 348, and NGC 7212).
This comparison suggests that there is no strong anti-correlation between
the presence of a hidden S1 nucleus and a nuclear starburst. This is at odds with
a model in which one class of S2 nuclei are powered by hidden S1 nuclei and another
class is powered by a nuclear starburst.

Our HST ultraviolet observations of Mrk 477, NGC 5135, NGC 7130 and IC 3639 have shown
that the starbursts in these S2 nuclei are very powerful, dusty and compact. They have
sizes that are similar to the NLR, and they are also similar to the torus sizes reported
through detections at near-IR and millimeter wavelengths (Young et al 1996;
Tacconi et al 1994). This similarity in the sizes suggests that these starbursts may be
formed at the outskirts of the molecular torus. This is a natural location where the star
formation can be triggered, because the torus may serve as a reservoir of molecular gas
at a few hundred parsecs from the nucleus. Thus, if the obscuring torus is a basic
component of the Seyfert phenomenon, nuclear starbursts should be ubiquitous in S2 nuclei.

If this is the case, why don't we see starbursts in all S2 nuclei? 
One possibility is that more sensitive spectroscopy (perhaps at the
much higher spatial resolution of HST+STIS) would reveal compact but inconspicuous starbursts
in the S2 nuclei whose ground-based spectra are dominated by the light
of old stars on the few-arcsec-scale. Alternatively, perhaps the obscuring torus sustains
intermittent bursts of star-formation with a duty cycle of roughly 50$\%$. The results reported
by GD98 are consistent with this picture, since they found that the stellar population
in four well-studied S2 nuclei was consistent with short-lived bursts with durations of
only a few Myr. On the other hand, such a picture would be inconsistent with the absence of
a significant intermediate age stellar population in roughly half our S2 sample (those
whose light is dominated by old stars).

\subsection{The nature of the blue continuum: breaking the young star versus power law degeneracy}

In order to know the nature of the FC in Seyfert nuclei we need to break the degeneracy 
between power-law and young stars. Our analysis has shown that the blue optical continuum is provided by a
nuclear starburst that dominates the near-ultraviolet light in about half of the S2 
in the sample. In these nuclei, the degeneracy between power-law and young
stars is completely broken because the blue continuum at the near-ultraviolet is not
feature-less at all; on the contrary, it is dominated by absorption features (HOBS and HeI lines) 
formed in the photosphere of young stars. Thus, the photospheric HOBS and HeI are a
powerful diagnostic to discriminate between a blue continuum povided by a young and intermediate age
population or by scattered light from a hidden S1. G-band and CaII H and K lines cannot help to break the
degeneracy because these lines can be diluted in the same way by young stars or by a scattered light.

However, if the starburst is in a nebular phase (younger than 
about a few Myr), Balmer absorption lines cannot help much to break the degeneracy. 
The stellar lines (the equivalent widths of the HOBL is less of 3 \AA) are
filled by the nebular emission provided by this young starburst (see figure 16 in Gonz\'alez Delgado,
Leitherer \& Heckamn 1999). HeI $\lambda$3819, 4026 and 4388 lines may be
a better diagnostic since after 4-5 Myr, this HeI nebular emission provided by the starburst 
drop and the stellar absorption dominates. In practice, it is difficult 
to break the degeneracy with HeI lines because these are weaker and they may be also filled by the
NLR contribution. The detection of WR features, as in the case Mrk 477, may also help to break the
degeneracy. The equivalent  width of the Balmer emission cannot be a good diagnostic because the NLR can
also contribute to the nebular emission, and the bulge stellar population can contribute to the  optical
continuum, and differential extinction can affect unevenly gas and  stars. Of course, in 
Seyfert nuclei with very young starbursts, the degeneracy is totally broken in the ultraviolet because O
stars are very rich in wind resonance lines (CIV, SiIV, NV, etc).

\section{Summary and conclusions}

Our most important conclusions from the near-ultraviolet and optical observations of
20 of the brightest known Seyfert 2 nuclei are as follows:

\begin{itemize}

\item{In 8 of the nuclei (Mrk 3, Mrk 34, Mrk 348, Mrk 573, NGC 1386, NGC 2110, NGC~5929
and NGC 7212) the stellar lines are similar to those of an old population. The presence of
scattered light produced by the hidden nucleus is required to fit the spectra in only four
of them, Mrk 3, Mrk 34, Mrk 573 and NGC 7212. However, only Mrk 3 shows a significant
dilution of the stellar lines in its nucleus compared to its circumnuclear region. The lack of
extra dilution in the nuclei of Mrk 34, Mrk 573, and NGC 7212 suggests the presence of a 
spatially-extended ``featureless''
continuum.}

\item{The spectral energy distribution and the strength of the stellar lines in the nuclei
and circumnuclear regions of Mrk 78 and NGC 1068 require a significant contribution of an intermediate
age population in addition to the old bulge component. However, 10\%\ and 15\%\ (respectively)
of the total optical light could also be due to scattered light from their
hidden Seyfert 1 nuclei. In the case of NGC 1068, the result is in agreement with the fact that
its optical continuum is highly polarized.}

\item{Six of the nuclei (Mrk 273, Mrk 1066, Mrk 1073, NGC 5135, NGC 7130 and IC 3639) show
HeI and/or strong Balmer absorption lines in the near-ultraviolet. Three other galaxies (Mrk 1, Mrk
477 and Mrk 533) show the higher order terms of the Balmer series in absorption in their
circumnuclear regions, and in their nuclear spectra when the nebular emission 
contributions are subtracted. Mrk 463E may be a similar object, but better data are needed to confirm this.
The analysis of the stellar lines using evolutionary
synthesis models suggests that the dominant stellar population in these nuclei
consists of young and intermediate age stars. These galaxies show a significant dilution of
the lines in their nuclei compared to their circumnuclear regions. Thus,
hot (young) stars are more conspicuous in the nucleus than in the bulge or inner disk.}

\item{A broad emission feature centered at around 4680 \AA\ is present in three nuclei
in our sample: Mrk 1, Mrk 463E, and Mrk 477. It is most plausibly ascribed to a population
of Wolf Rayet stars (cf. H97), but confirmation will require the detection of other
signatures of Wolf Rayet stars.}

\item{Comparison 
with the prototypical starburst nucleus in NGC 7714 indicates that roughly half
of the Seyfert 2 nuclei in our sample show 
characteristics that are very similar to starburst nuclei. Therefore, it is
expected that young massive stars will dominate the ultraviolet spectra of these nuclei.
This has been directly confirmed for four members of our sample (H97, GD98)}.

\item{The nuclei with young and intermediate-age populations show lower excitation
(smaller [OIII]/H$\beta$ flux ratio), larger far-IR luminosities, and
cooler f$_{25\mu}$/f$_{60\mu}$ IRAS colors than nuclei dominated by an old stellar population.
These differences are as expected if a starburst makes a substantial contribution
to the heating of the dust and the ionization of the gas in the former set of nuclei,
but not the latter.}

\item{The fact that hidden BLRs have been found in nuclei with young and
intermediate-age stars (Mrk 477, Mrk 533 and IC 3639) implies that the presence of starbursts 
and hidden Seyfert 1 nuclei are not
anti-correlated. That is, there is no evidence for two distinct types of Seyfert 2 nuclei
(starburst-powered and AGN-powered).} 

\end{itemize}  

{\bf Acknowledgments}

We are grateful to Enrique P\'erez and Luis Colina for many stimulating
discussions and
helpful suggestions, and to the staff at KPNO for their help 
in obtaining the data presented in this paper. This work was supported by
Spanish DGICYT projects PB98-0521, by the NASA LTSA grant NAG5-6400,
and by HST grants GO-5944 and GO-6539
from the Space Telescope Science Institute, which is operated by the
Association of Universities for Research in Astronomy, Inc., under NASA
contract NAS5-26555.

\clearpage

\begin{deluxetable}{lcccccccc}
\footnotesize
\tablecaption{Properties of the Seyfert 2 galaxies \tablenotemark{a}}
\tablewidth{0pt}
\tablehead{
\colhead{Name} & \colhead{Type} & \colhead{v} & \colhead{1 arcsec} & \colhead{log L$_{[OIII]}$} & 
\colhead{log L$_{1.4}$} &  \colhead{log L$_{12}$} &  \colhead{L$_{FIR}$} & \colhead{L$_{IR}$} \nl
\colhead{} & \colhead{} & \colhead{km/s} & \colhead{pc} & \colhead{erg s$^{-1}$} & \colhead{erg s$^{-1}$} 
& \colhead{L$\odot$} & \colhead{10$^{10}$ L$\odot$} & \colhead{10$^{11}$ L$\odot$}
} 
\startdata

 Mrk 1     & S             & 4780  & 310 & 41.62 & 38.69  & 10.77 & 1.5 & 0.9 \nl
 Mrk 3     & S0:           & 4050  & 260 & 42.28 & 39.78  & 10.20 & 1.5 & 0.6 \nl
 Mrk 34    & S:            & 15150 & 980 & 42.68 & 39.09  & 10.33 & 4.6 & 1.4 \nl
 Mrk 78    & SB            & 11145 & 720 & 42.39 & 39.11  & 10.34 & 3.6 & 1.1 \nl
 Mrk 273   & Ring Gal. pec & 11334 & 730 & 41.97 & 39.73  & 10.61 & 71.0& 12.0\nl
 Mrk 348   & SA(s)0/a:     & 4540  & 290 & 41.46 & 39.39  & 10.15  & 0.7 & 0.3 \nl
 Mrk 463E  & S pec         & 14904 & 960 & 42.70 & 40.46  & 11.19 & 11.7& 5.0 \nl
 Mrk 477   & S             & 11340 & 730 & 42.79 & 39.40  & 10.34 & 5.5 & 1.5 \nl
 Mrk 533   & SA(r)bc pec   & 8713  & 563 & 42.03 & 39.70  & 10.85 & 11.9& 3.1 \nl
 Mrk 573   & SAB(rs)0+:    & 5174  & 335 & 42.07 & 38.03  & 9.96  & 0.8 & 0.3 \nl
 Mrk 1066  & SB(s)0+       & 3605  & 235 & 40.97 & 38.58  & 9.90  & 3.6 & 0.6 \nl
 Mrk 1073  & SB(s)b        & 6991  & 450 & 41.50 & 39.35  & 10.47 & 10.9& 2.2 \nl
 NGC 1068  & SA(rs)b       & 1136  & 75  & 41.88 & 39.17  & 10.85 & 6.1 & 2.1 \nl
 NGC 1386  & SB(s)0+       & 924   & 60  & 40.64 & 37.12  & 8.76  & 0.1 & 0.03 \nl
 NGC 2110  & SAB0-         & 2284  & 150 & 40.44 & 38.63  & 9.40  & 0.6 & 0.1 \nl
 NGC 5135  & SBab          & 4112  & 270 & 41.05 & 38.93  & 10.17 & 3.9 & 0.9 \nl
 NGC 5929  & Sab:pec       & 2753  & 180 & 40.40 & 38.36 & 9.65  & 2.0 & 0.4 \nl
 NGC 7130  & Sa pec        & 4842  & 310 & 41.28 & 38.99  & 10.28 & 11.1 & 2.0 \nl
 NGC 7212  & S             & 7994  & 515 & 42.09 & --     & 10.23 & 5.5 &  1.2\nl
 IC 3639   & SBbc          & 3285  & 210 & 40.77 & 38.47  & 10.00  & 2.3 & 0.5 \nl

\enddata
\tablenotetext{a} {Data for the [OIII] $\lambda$5007+4959 and 1.4 GHz radio continuum are from Whittle (1992b). 
The FIR luminosity was computed from the 60 and 100 $\mu$m IRAS fluxes (Fullmer \& Lonsdale 1989); and IR luminosity 
was computed using all 4 IRAS fluxes (Sanders \& Mirabel 1997). Radial velocities and morphological type are from
NED. 
H$_0$=75 km s$^{-1}$ Mpc$^{-1}$}
\end{deluxetable}

\begin{deluxetable}{lcccccccc}
\footnotesize
\tablecaption{Log of the Observations}
\tablewidth{0pt}
\tablehead{
\colhead{Name} & \colhead{Date} & \colhead{Exp. Time} & \colhead{P.A.} & \colhead{Air mass} & 
\colhead{Date} & \colhead{Exp. Time} & \colhead{P.A.} & \colhead{Air mass} \nl
\colhead{} & \colhead{} & \colhead{s} & \colhead{degree} & \colhead{} & 
\colhead{} & \colhead{s} & \colhead{degree} & \colhead{} \nl
} 
\startdata

 Mrk 1     & Oct 11  & 2$\times$1200  & 95  & 1.02 & Oct 11 & 2$\times$1800 & 95  & 1.0   \nl    
 Mrk 3     & Feb 16  & 1200, 600      & 164 & 1.3  & Feb 15 & 2$\times$1800 & 144 & 1.3   \nl
 Mrk 34    & Feb 16  & 900, 1200      & 37  & 1.16 & Feb 14 & 2$\times$1800 & 162 & 1.15  \nl
 Mrk 78    & Feb 16  & 2$\times$900   & 165 & 1.2  & Feb 15 & 2$\times$1800 & 160 & 1.2   \nl
 Mrk 273   & Feb 16  & 900,1200       & 73  & 1.3  & Feb 15 & 3$\times$1800 & 70  & 1.2   \nl
 Mrk 348   & Oct 10  & 2$\times$1200  & 98  & 1.0  & Oct 11 & 2$\times$1800 & 98  & 1.05    \nl
 Mrk 463E  & Feb 16  & 900,600        & 160 & 1.0  & Feb 14 & 2$\times$1800 &175,134& 1.05  \nl
 Mrk 477   & Feb 16  & 2$\times$1200  &144,158& 1.0& Feb 14 &1800,900       & 0   & 1.1   \nl
 Mrk 533   & Oct 10  & 2$\times$1200  & 162 & 1.1  & --     & --            & --  & --    \nl
 Mrk 573   & Oct 10  & 2$\times$1200  & 161 & 1.15 & Oct 11 & 2$\times$1800 & 161 & 1.2      \nl
 Mrk 1066  & Oct 10  & 2$\times$1200  & 96  & 1.1  & Oct 11 & 1800,900      & 96  & 1.1      \nl
 Mrk 1073  & Oct 10  & 1200,900       & 96  & 1.2  & Oct 11 & 1800,900      & 96  & 1.2    \nl
 NGC 1068  & Oct 10  & 900,600,300    & 123 & 1.2  & Oct 11 & 2$\times$900  & 123 & 1.2     \nl
 NGC 1386  & Oct 10  & 2$\times$1200  & 360 & 2.6  & Oct 11 & 2$\times$1200 & 0   & 2.6   \nl
 NGC 2110  & Feb 16  & 2$\times$900   & 6   & 1.3  & Feb 15 & 1800          & 5   & 1.3   \nl
 NGC 5135  & Feb 16  & 2$\times$900   & 173 & 2.1  & --     & --            & --  & --    \nl
 NGC 5929  & Feb 16  & 2$\times$900   &60,80& 1.03 & Feb 15 & 1800,900      & 80  & 1.1     \nl
 NGC 7130  & Oct 10  & 1200,1800      &177  & 2.55 & Oct 11 & 1800          & 177 & 2.55   \nl
 NGC7212   & Oct 10  & 2$\times$1200  & 136 & 1.15 & Oct 11 & 2$\times$1800 & 136 & 1.1    \nl
 IC 3639   & Feb 16  & 2$\times$1200  & 170 & 2.7  & --     & --            & --  & --    \nl
 NGC 205   & Feb 16  &300,2$\times$600& 82  & 1.6  & Feb 15 & 300,2$\times$600&85 & 1.55 \nl
 NGC 1569  & Feb 16  & 2$\times$600   & 140 & 1.2  & Feb 14 & 2$\times$600  & 140 & 1.2   \nl
 NGC 4339  & Feb 16  & 600,900        & 130 & 1.35 & Feb 15 &300,2$\times$600&135 & 1.2  \nl

\enddata

\end{deluxetable}

\begin{deluxetable}{lcccccc}
\footnotesize
\tablecaption{Equivalent widths (in \AA\ units) of the H$\beta$ emission-line and the stellar absorption
lines  CaII K $\lambda$3918, G band and the Calcium triplet at 
$\lambda$8498, $\lambda$8542, and $\lambda$8662}
\tablewidth{0pt}
\tablehead{
\colhead{Name} & \colhead{H$\beta$ emiss} & \colhead{CaII K} & \colhead{G band} & \colhead{CaII $\lambda$8498} 
& \colhead{CaII $\lambda$8542} & \colhead{CaII $\lambda$8662} \nl
} 
\startdata

 Mrk 1     & 69$\pm$3     & 5.6$\pm$0.5 & 3.1$\pm$0.3 & 1.3$\pm$0.3 & 2.8$\pm$0.3 & 1.9$\pm$0.3          \nl
 Mrk 3     & 73$\pm$3     & 8$\pm$1.3   & 4.1$\pm$0.2 & 1.4$\pm$0.2 & 2.4$\pm$0.2 & 1.6$\pm$0.2           \nl
 Mrk 34    & 45$\pm$2     & 8$\pm$1.0   & 4.8$\pm$0.4 & -   & -     &  -            \nl
 Mrk 78    & 12.5$\pm$0.5 & 8.3$\pm$0.9 & 3.9$\pm$0.3 & 1.4$\pm$0.1 & 3.5$\pm$0.1 & -              \nl
 Mrk 273   & 11.4$\pm$0.6 & 5.0$\pm$0.5 & 1.9$\pm$0.2 & -           & 2.6$\pm$0.3 & -             \nl
 Mrk 348   & 23.3$\pm$0.8 & 9.5$\pm$1.2 & 5.9$\pm$0.5 & 1.3$\pm$0.2 & 3.2$\pm$0.2 & 3.3$\pm$0.2           \nl
 Mrk 463E  & 118$\pm$6    & 0.9$\pm$0.7 & 0   & -   & -   & -             \nl
 Mrk 477   & 126$\pm$5    & 0.8$\pm$0.3 & 0.6$\pm$0.3 & -           & 2.6$\pm$0.4 & -             \nl
 Mrk 533   & 45$\pm$1.4   & 3.8$\pm$0.4 & 1.9$\pm$0.3 & -           & -   & -             \nl
 Mrk 573   & 22.0$\pm$0.9 & 11$\pm$1.5  & 5.5$\pm$0.6 & 1.3$\pm$0.2 & 3.5$\pm$0.2 & 3.1$\pm$0.2           \nl
 Mrk 1066  & 25.7$\pm$0.6 & 4.8$\pm$0.3 & 1.8$\pm$0.3 & 1.3$\pm$0.1 & 2.9$\pm$0.1 & -             \nl
 Mrk 1073  & 21.2$\pm$0.5 & 5.0$\pm$0.4 & 2.6$\pm$0.3 & 1.0$\pm$0.2 & 2.7$\pm$0.3 & -             \nl
 NGC 1068  & 39.0$\pm$1.3 & 6.3$\pm$0.5 & 2.7$\pm$0.3 & 1.2$\pm$0.1 & 2.9$\pm$0.1 & 2.6$\pm$0.1           \nl
 NGC 1386  & 3.2$\pm$0.4  & 12.2$\pm$0.8& 5.9$\pm$0.6 & 2.2$\pm$0.1 & 3.8$\pm$0.1 & 3.3$\pm$0.1           \nl
 NGC 2110  & 15.8$\pm$0.9 & 11$\pm$1.5  & 5.6$\pm$0.3 & 1.3$\pm$0.1 & 2.8$\pm$0.1 & 2.1$\pm$0.1           \nl
 NGC 5135  & 13.8$\pm$0.3 & 3.1$\pm$0.2 & 1.8$\pm$0.3 & -   & -   & -            \nl
 NGC 5929  & 14.0$\pm$0.7 & 11$\pm$1.0  & 7.1$\pm$0.7 & 1.3$\pm$0.2 & 3.0$\pm$0.2 & 2.2$\pm$0.2          \nl
 NGC 7130  & 17.4$\pm$0.4 & 3.0$\pm$0.2 & 1.6$\pm$0.4 & 1.3$\pm$0.2 & 3.0$\pm$0.2 & -            \nl
 NGC7212   & 60$\pm$3     & 8.3$\pm$0.9 & 4.3$\pm$0.4 & 1.2$\pm$0.6 & 2.1$\pm$0.6 & -            \nl
 IC 3639   & 18.5$\pm$0.7 & 6.8$\pm$0.4 & 4.1$\pm$0.4 &  -  & -   & -    \nl
 NGC 205   & -            & 5.6$\pm$0.6 & 3.9$\pm$0.5 & 1.2$\pm$0.2 & 2.8$\pm$0.2 & 2.4$\pm$0.2 \nl 
 NGC 1569B & -            & 1.8$\pm$0.2 & 1.5$\pm$0.2 & -           & 3.7$\pm$0.3 & 3.2$\pm$0.1  \nl
 NGC 4339  & -            & 14$\pm$1.4  & 8$\pm$1.1   & 1.3$\pm$0.1 & 3.6$\pm$0.1 & 3.1$\pm$0.1 \nl

\enddata

\end{deluxetable}

%
%

\clearpage


\figcaption{Equivalent width of the H$\beta$ line in emission versus the
equivalent width of the stellar G band,  measured in the nuclear spectra corresponding 
to a 1.5$\times$2.1
arcsec$^2$ aperture centered at the peak of the optical continuum at
4550 \AA.
The different symbols correspond to the four groups in which the objects
are classified.
Objects with only very weak stellar absorption-lines (Mrk 463E and Mrk 477) are
represented by diamonds;
those with a continuum dominated by an old stellar population (Mrk 3,
Mrk 34,
Mrk 348, Mrk 573, NGC 1386, NGC 2110, NGC 5929 and NGC 7212) by
triangles; 
objects with a dominant young and intermediate-age stellar
population (Mrk 273, Mrk 1066,  Mrk 1073, NGC 5135, NGC 7130, IC 3639,
Mrk 1 and Mrk 533) by open circles; and
objects in which the stellar population has an important contribution from
intermediate-age stars (Mrk 78
and NGC 1068) by filled circles.  Dashed lines mark the values of the Ew(G
band) of the post-starburst nucleus of NGC 205 (3.6
\AA), of the typical elliptical galaxy NGC 4339 (7.6 \AA), and of the super-star cluster B in 
the starburst NGC
1569 (1.5 \AA).}

 
\figcaption{Normalized off-nuclear spectrum of Mrk 477 obtained through
a 1.5$\times$2
arcsec$^2$ aperture at 2 arcsec from the optical continuum peak.
The nuclear spectrum of the companion galaxy of Mrk 477 is shown for
comparison (dotted line).
Both show the higher order terms of the Balmer series in
absorption. }



\figcaption{Ground-based optical spectrum of Mrk 463E obtained through a
1.5$\times$2.1
arcsec$^2$ aperture. It shows a broad emission feature at 4680 \AA\
that may be due to a population of Wolf Rayet stars.}



\figcaption{{\it HST\/}+FOC (through F210M) image of the central
2$\times$2 arcsec$^2$ (2 by 2
kpc) of Mrk 463E. The morphology is different from the other S2 galaxies
that  harbor a nuclear
starburst. The S2 nucleus is located at the brightest UV knot, whose nature
is unknown. Most of the UV flux comes from 1 kpc-scale extended emission.}


 

\figcaption{Nuclear spectra of three objects NGC 7212 (a), NGC 5929 (b),
and NGC 2110 (c and d)
that belong to the second group, i.e., nuclei with a stellar population
similar to that of
elliptical  galaxies. Also shown is the best fit to the optical
continuum, obtained by combining
the elliptical galaxy NGC 4339 plus a power-law (a, b and c). The
power-law accounts for 15$\%$,
10$\%$ and 5$\%$ of the total light at 4800 \AA\ in NGC 7212, NGC 5929
and NGC 2110,
respectively. The nebular Balmer continuum is also added to the fit of
NGC 7212, contributing
5$\%$ of the total light at 4800 \AA. Figure 5d shows the result of
fitting the
nuclear spectrum of NGC 2110 with a circumnuclear spectrum that is
reddened by
\mbox{E(B-V)=0.2}. This spectrum is the average of two spectra extracted at 5
arcsec from the peak of
the optical continuum at 4550 \AA. The spectra are all normalized to the
flux of the continuum
at 4800 \AA.}



\figcaption{Nuclear spectra of objects that show the higher order terms
of the Balmer series in
absorption (Mrk 273, Mrk 1066, Mrk 1073, NGC 5135, NGC 7130, and IC
3639). All the spectra are
normalized to the continuum flux at 4800 \AA. They are plotted with an
arbitrary offset in flux for clarity.}



\figcaption{The nuclear spectrum of Mrk 1066 normalized to the flux at
4800 \AA, and the fit
resulting from combining a circumnuclear spectrum plus a power-law that
contributes 30$\%$ to
the total light at 4800 \AA\ (dotted line). The circumnuclear spectrum
is extracted at 3.5
arcsec from the optical continuum peak.}



\figcaption{The nuclear spectrum of Mrk 533, after the nebular emission
of H$\delta$-H10 is
subtracted from the observed spectrum, is compared with an off-nuclear
spectrum (dotted line)
of Mrk 533 extracted 5 arcsec away from the optical continuum peak. Both show
similar
Balmer absorption features below 4000 \AA.}




\figcaption{As in Figure 8, but for Mrk 1. The circumnuclear spectrum (dotted
line) is extracted at 3
arcsec from the optical continuum peak. Both show similar Balmer absorption
features below 4000 \AA.}



\figcaption{ Ground-based optical spectrum of Mrk 1 after the absorption-lines due
 to old stars have 
been subtracted using the spectrum of a K giant star. It shows a broad emission feature 
at 4680 \AA\
that may be due to a population of Wolf Rayet stars.}



\figcaption{Nuclear spectrum of Mrk 78 normalized to the flux at 4800
\AA, compared with the
result of combining the elliptical galaxy NGC 4339 with the synthetic
spectrum predicted by our
evolutionary models for a 1 Gyr old instantaneous burst (dotted line).
Each component contributes
50$\%$ of the total light at 4800 \AA. The resulting spectrum fits
the stellar
features and the continuum spectral energy distribution. }



\figcaption{Nuclear spectrum of NGC 1068 compared with the result of
combining a
circumnuclear spectrum (extracted 4 arcsec away from the continuum peak) and
a power-law that
contributes 10$\%$ of the total light at 4800 \AA. The
nebular continuum
contributes 3$\%$ to the total light at 4800 \AA.}



\figcaption{Spatial variation of the G band equivalent width (full line)
and the
CaII K lines (dashed line) toward the nucleus of Mrk 477.
Both stellar lines show a significant dilution toward the nucleus. }



\figcaption{As in Figure 13, but for objects with a dominant old stellar population (NGC 1386, NGC
5929,  Mrk 573,
and Mrk 3). Note that the stellar lines are not significantly diluted in
the nuclei of these
objects, except in Mrk 3.  The two lines (for each stellar feature) in
NGC 5929 represent the
values measured in two different orientations of the slit.}



\figcaption{As in Figure 13, but for objects that show the higher order terms of
the Balmer series
in absorption (NGC 5135, Mrk 1066, Mrk 533 and IC 3639). The stellar
lines are
significantly diluted in the nuclei of these galaxies, probably due to
the contribution of a
young  stellar burst that is more conspicuous in the nucleus.}



\figcaption{As in Figure 13, but for NGC 1068. Note that the decrease of the
equivalent width of the
stellar lines at $\pm$ 10 arcsec away from the continuum peak is due to a
change in the stellar
population in NGC 1068, produced by the circumnuclear ring
of HII regions. The
dilution of the stellar lines in the nucleus is associated with the
contribution of scattered
light from the hidden Seyfert 1 nucleus.}



\figcaption{Nuclear spectrum of the post-starburst NGC 205 normalized to the flux at 4800
\AA, and the best fit
(dotted line) obtained by combining the spectrum of the elliptical
galaxy NGC 4339 with the
average spectrum predicted by our evolutionary models of an intermediate age post-burst
(modeled for convenience as a sum of instantaneous bursts of ages 200, 500 and
1000 Myr). The intermediate age population contributes 70$\%$ of the
total light at 4800
\AA. Note that the resulting spectrum fits the stellar lines and the
shape of the optical
continuum distribution.}



\figcaption{Normalized nuclear spectra of NGC 5135 (a), Mrk 1066 (b),
Mrk 273 (c), and IC 3639
(d), and the composite model (dotted line) that fits the stellar lines.
The best fits are
obtained combining the young burst (3 Myr old), an intermediate age
population (the average
spectrum of three instantaneous bursts of 200, 500 and 1000 Myr old) and
an old (10 Gyr) population.
The relative contributions to the total light at 4800 \AA\ are: (a)
50$\%$, 40$\%$ and 10$\%$;
(b) 30$\%$, 50$\%$ and 20$\%$; (c) 30$\%$, 40$\%$ and 30$\%$; (d)
20$\%$, 40$\%$ and 40$\%$,
respectively.}



\figcaption{As in Figure 17, but for the circumnuclear regions of:
 (a) Mrk 1 (extracted 3 arcsec from the nucleus); (b) NGC 1068 (``bulge''
spectrum extracted 4
arcsec from the continuum peak at 4550 \AA); (c) NGC 1068 (HII ring
extracted 10 arcsec
from the continuum peak);  (d) Mrk 477 (extracted 2 arcsec from the
nucleus). The best
fits  are obtained combining a young burst, an intermediate age
population, and an old 
population, except in the ``bulge'' spectrum of NGC 1068 (which combines
an intermediate age and
old stellar population).  The relative contributions to the total
light at 4800 \AA\ are: (a)
20$\%$, 40$\%$ and 40$\%$;  (c) 40$\%$, 30$\%$ and 30$\%$; (d) 30$\%$,
40$\%$ and 30$\%$;
for the young, intermediate and old stellar population, respectively;
and b) 50$\%$ and 50$\%$
for the intermediate and the old stellar population, respectively. }



\figcaption{Nuclear spectrum in the CaII triplet region normalized to
the continuum flux at 8400
\AA\ in (a) objects with a dominant old stellar population: Mrk 3, Mrk 573, NGC
1386 and NGC 2110, and
(b) objects that show the higher order terms of the Balmer series in
absorption: Mrk 1, Mrk 1066,
Mrk 1073 and NGC 7130. The nuclear spectra of the elliptical galaxy NGC
4339 and the post-starburst galaxy NGC 205
are shown for comparison.  }



\figcaption{Infrared colors f$_{25\mu}$/f$_{60\mu}$ versus the nuclear excitation ratio
[OIII]/H$\beta$. Symbols are as in Figure 1, and they are marked by the names of the
objects. }


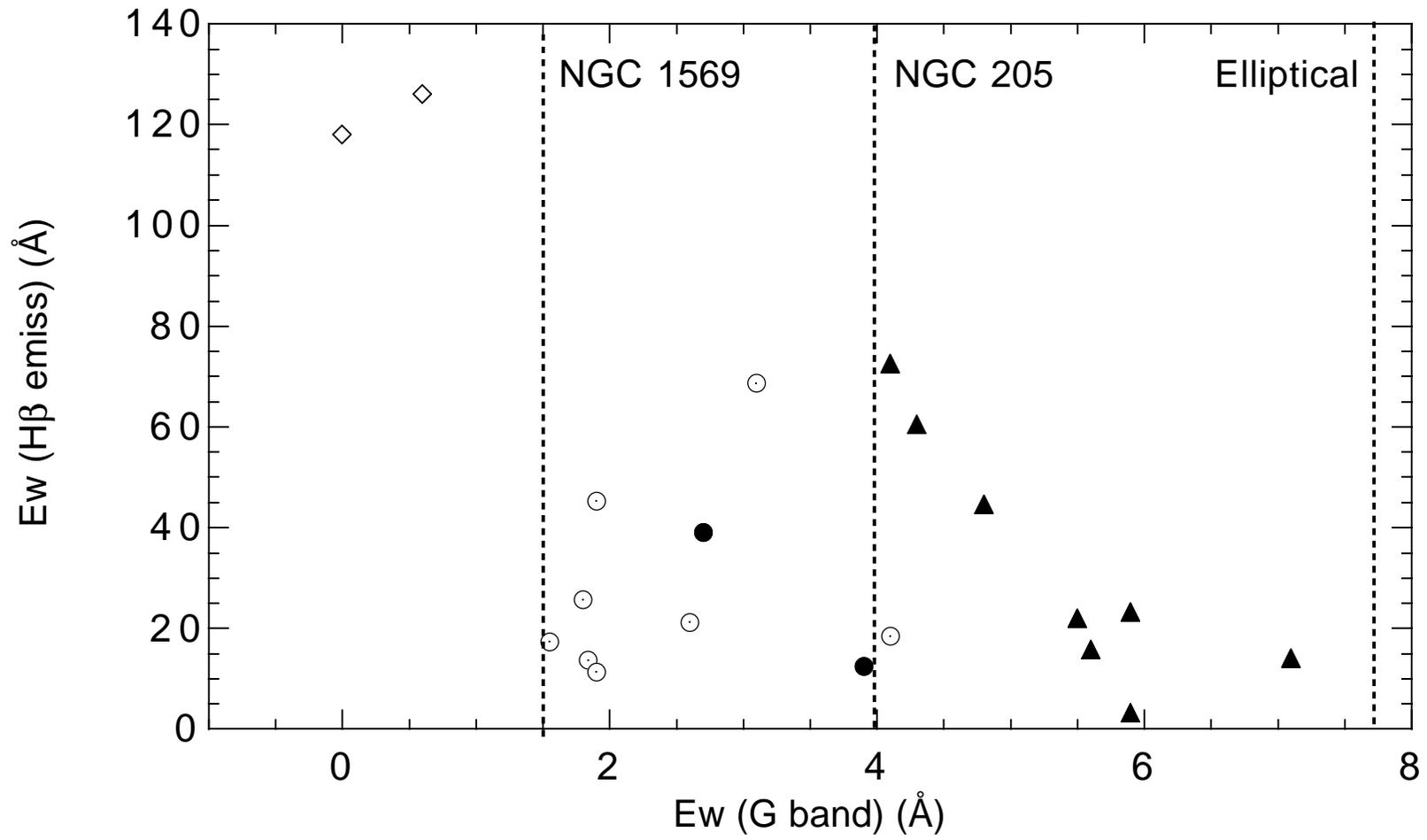
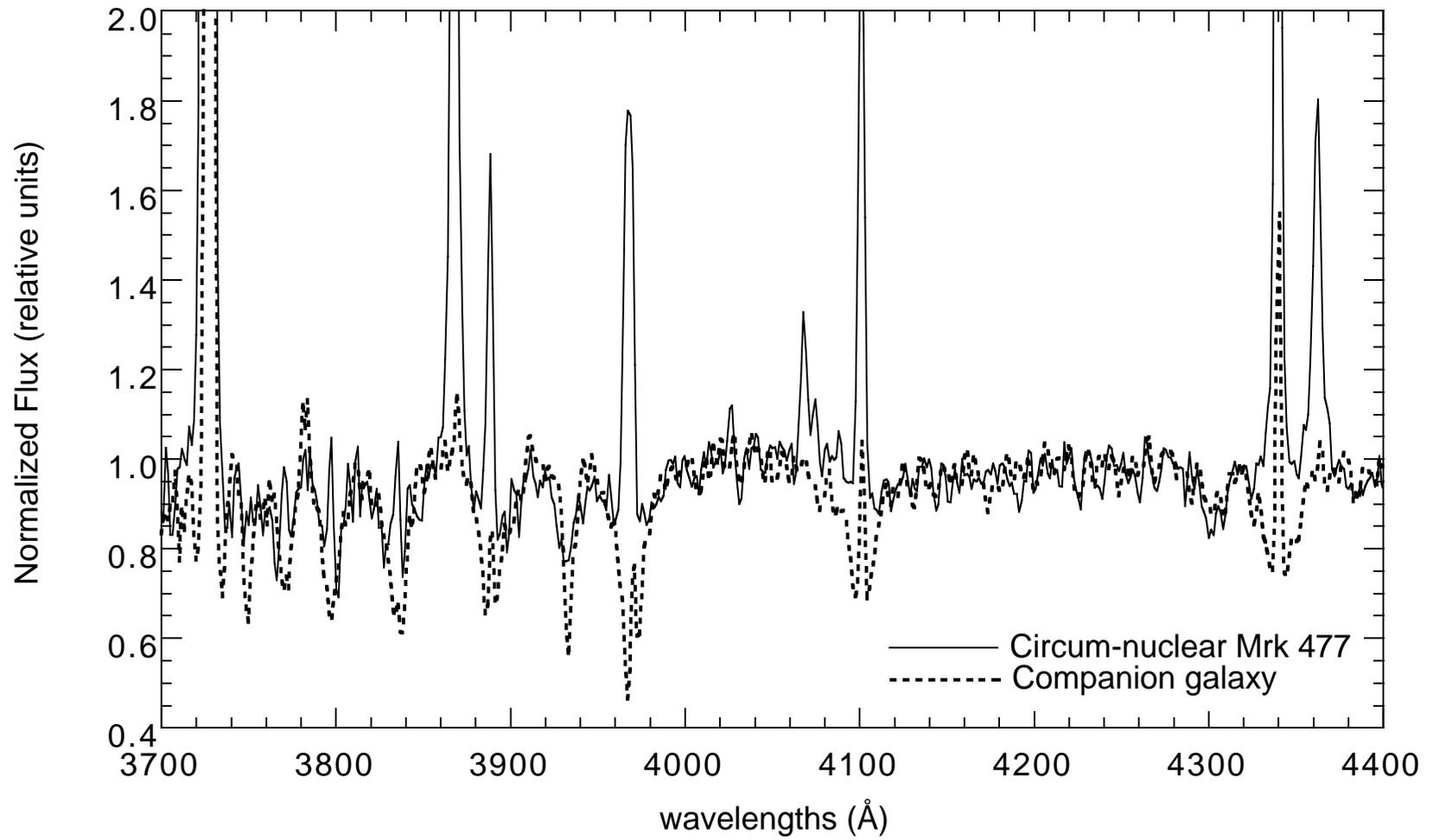
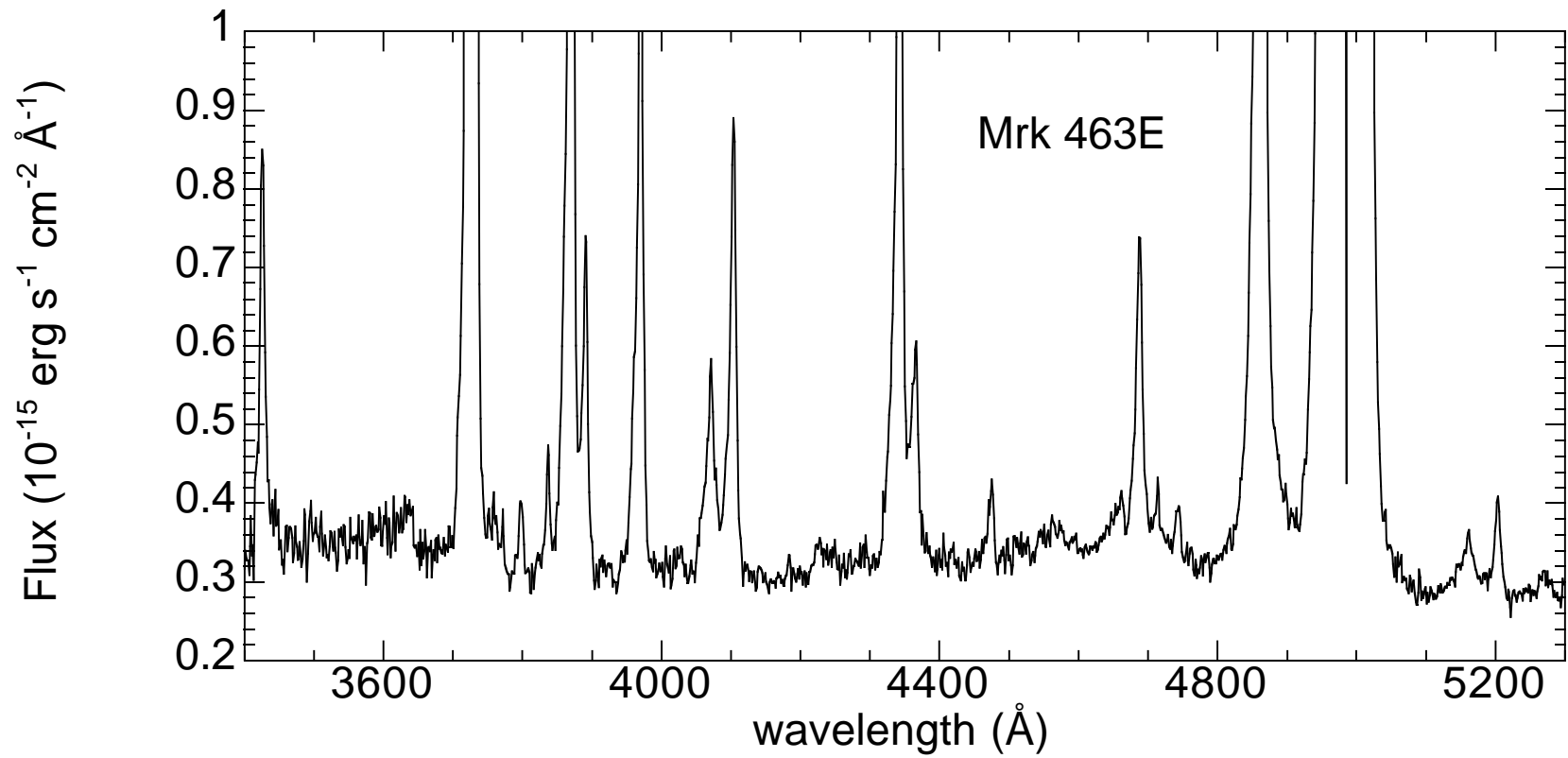
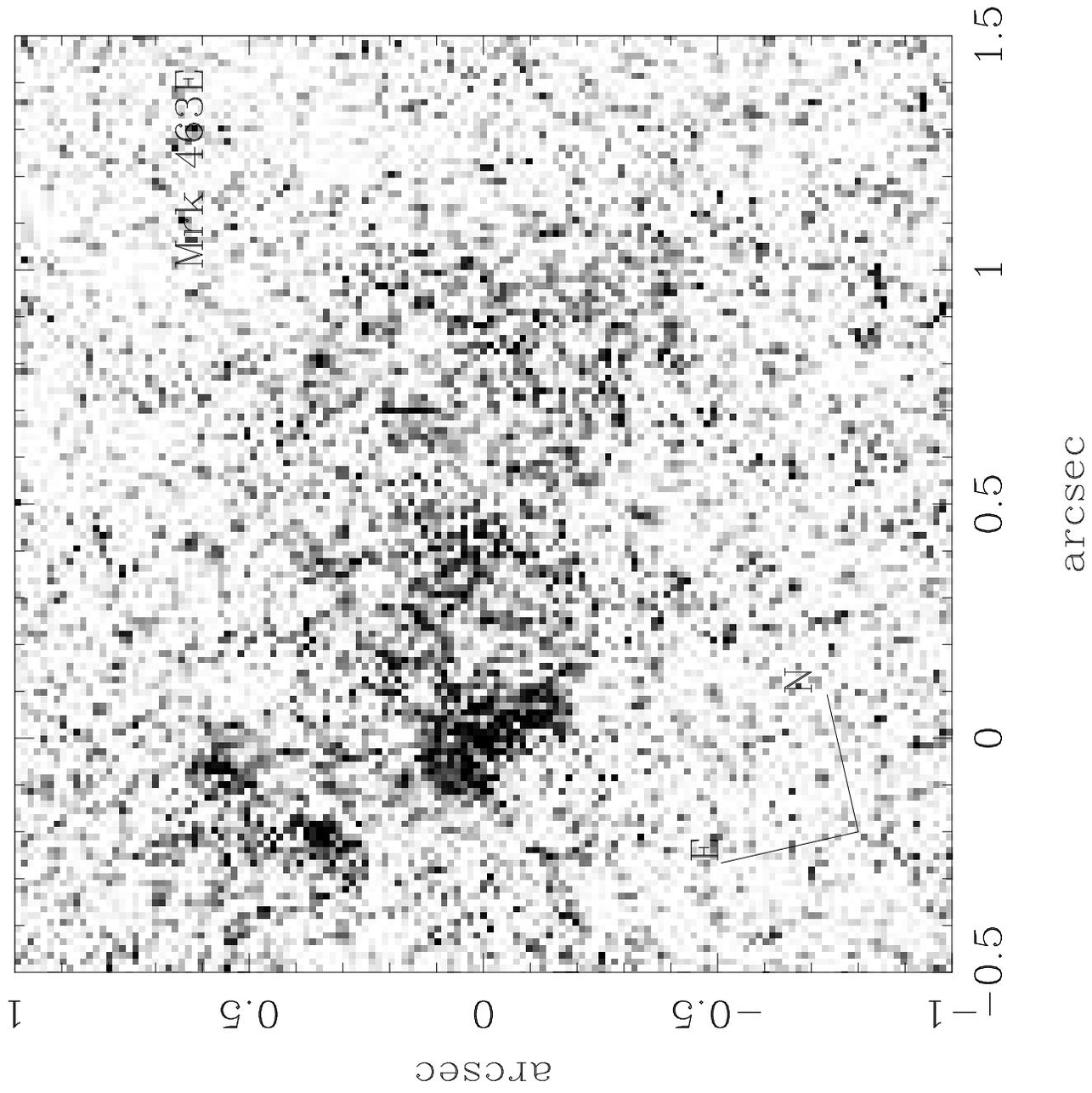
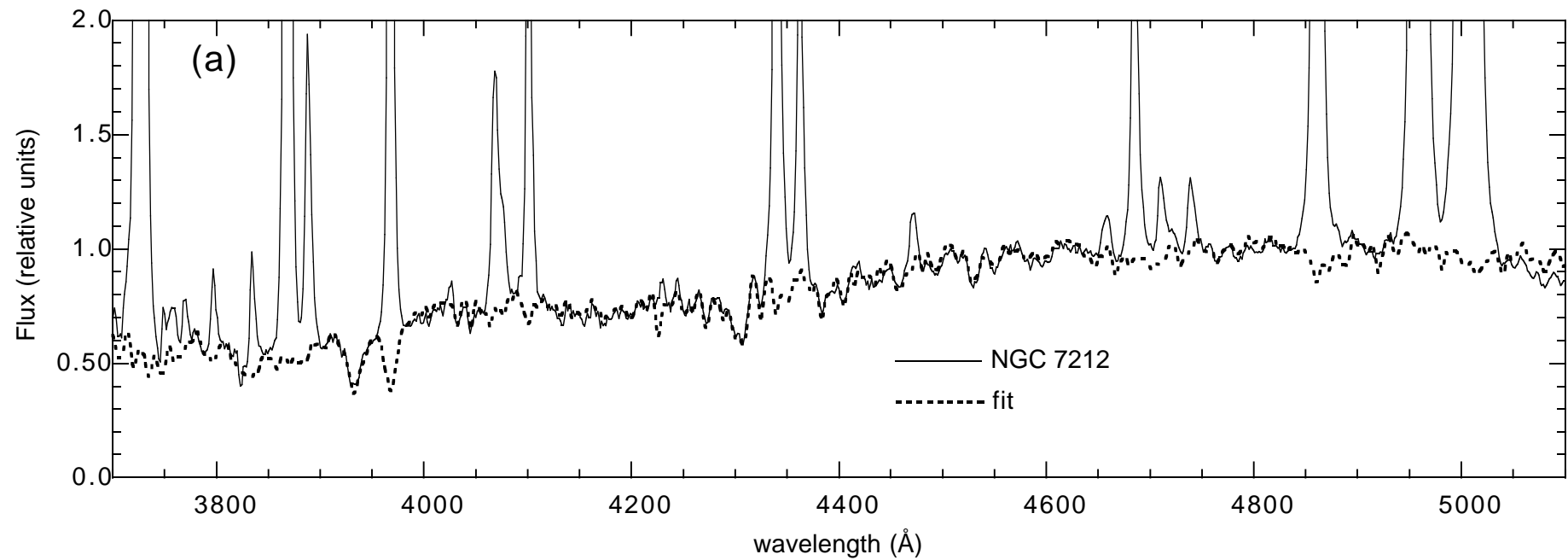
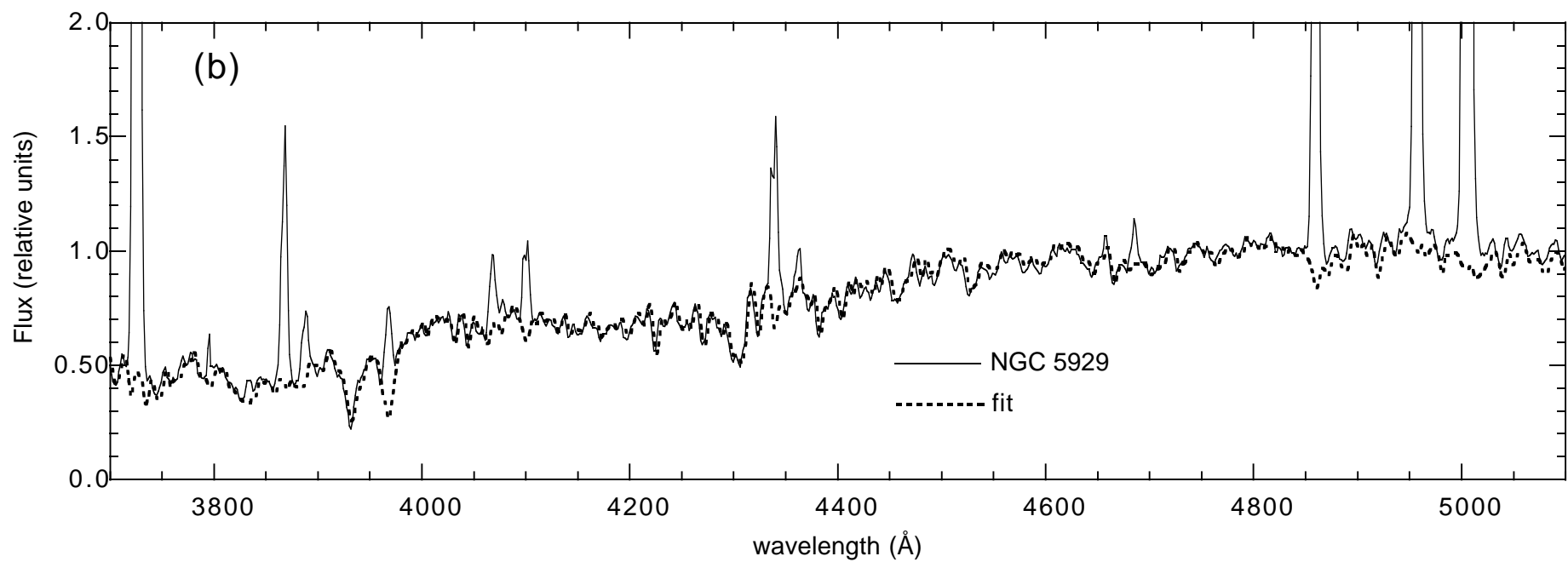
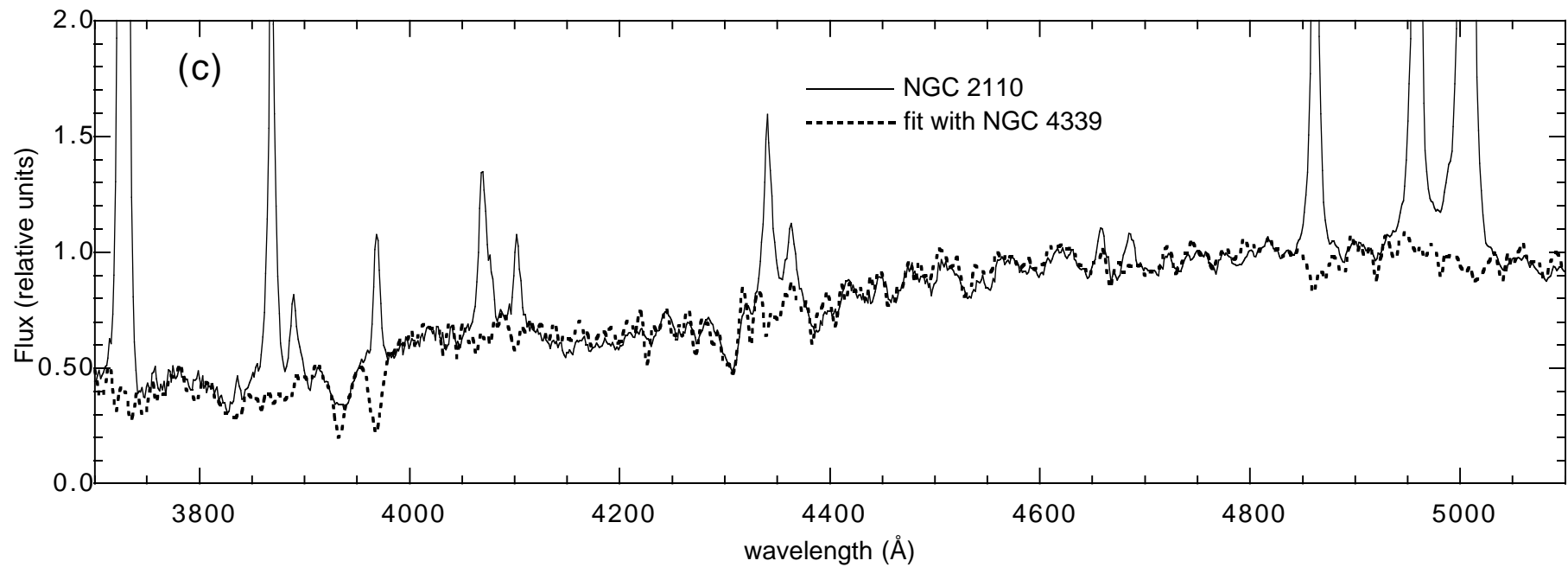
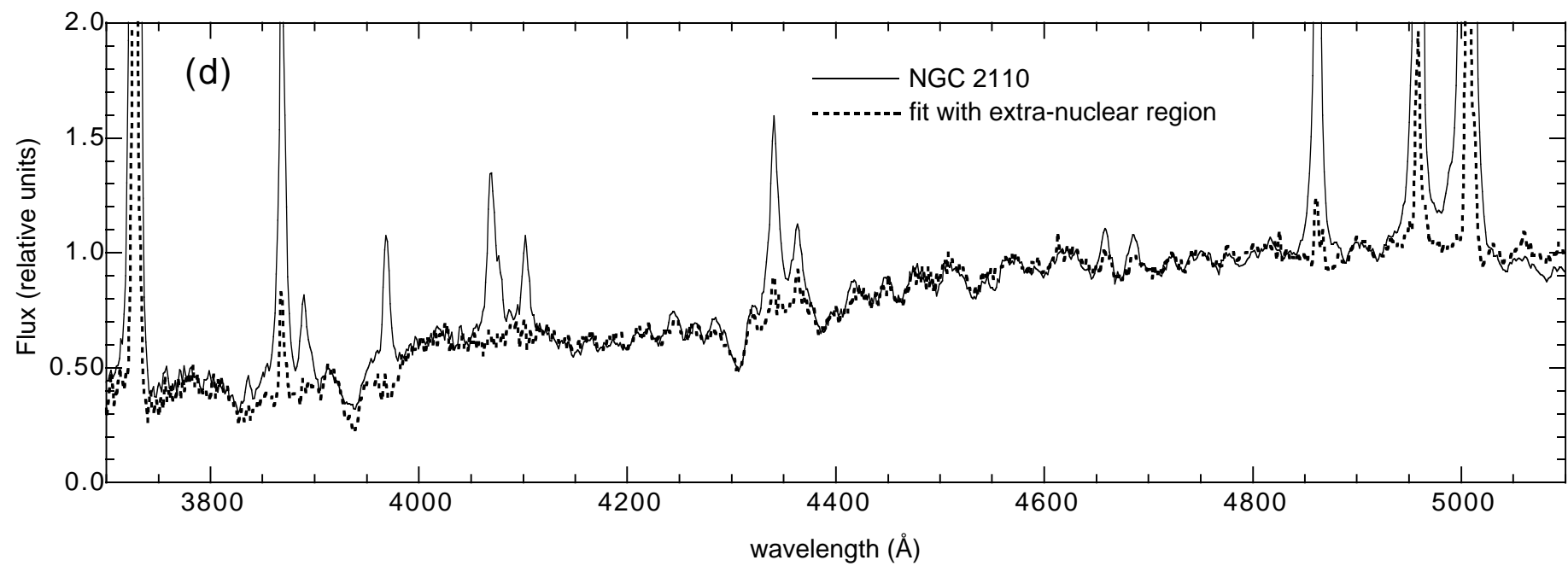
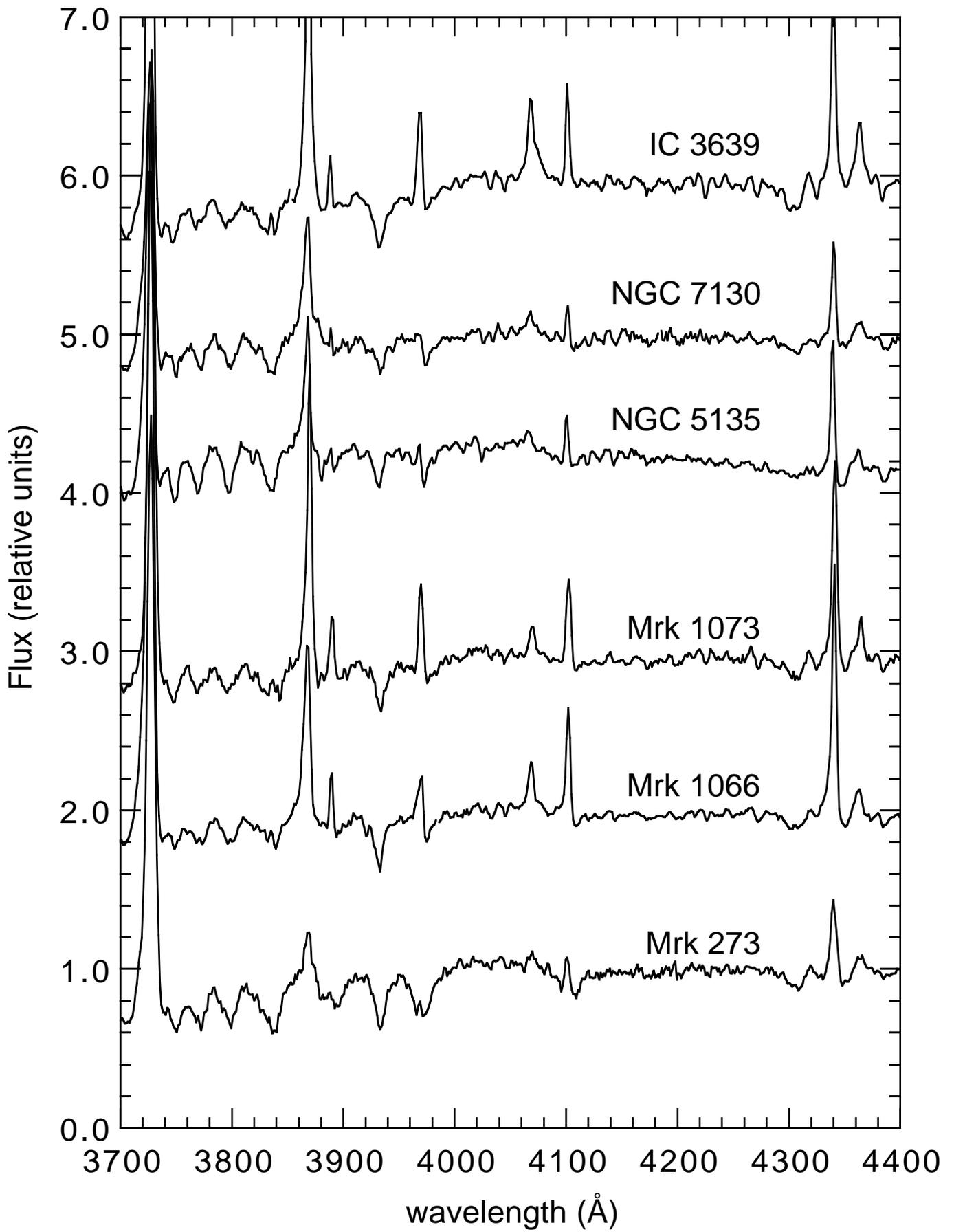
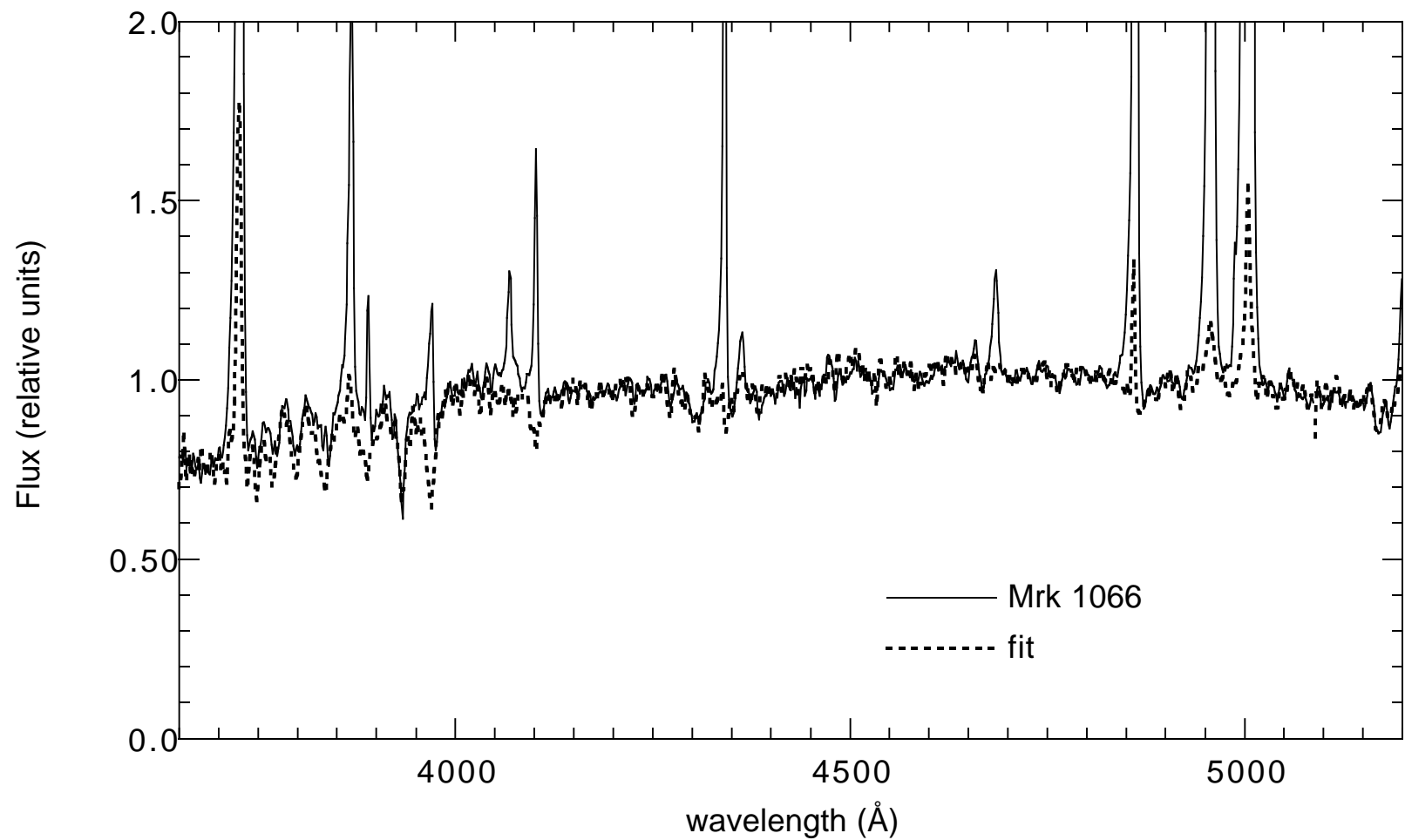
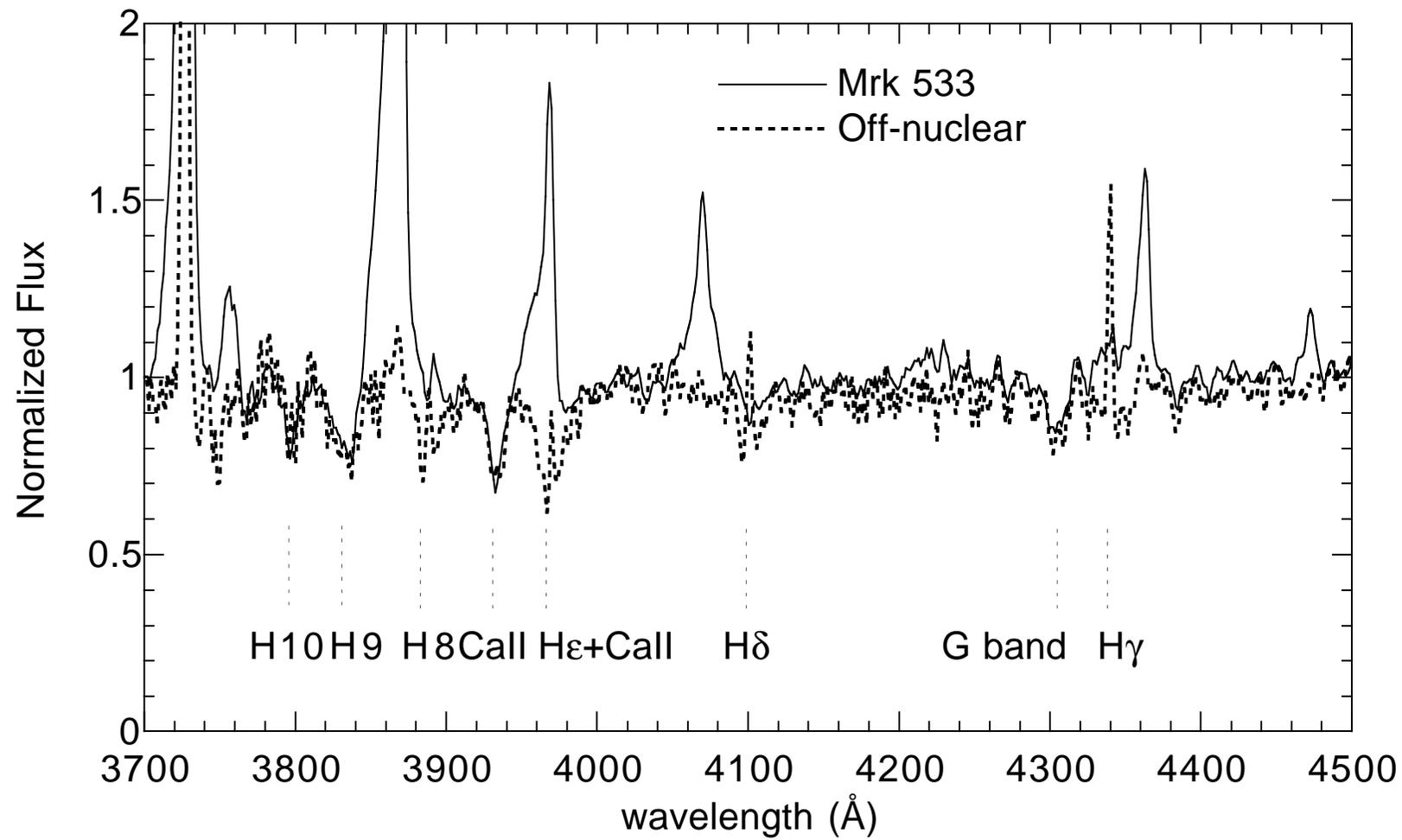
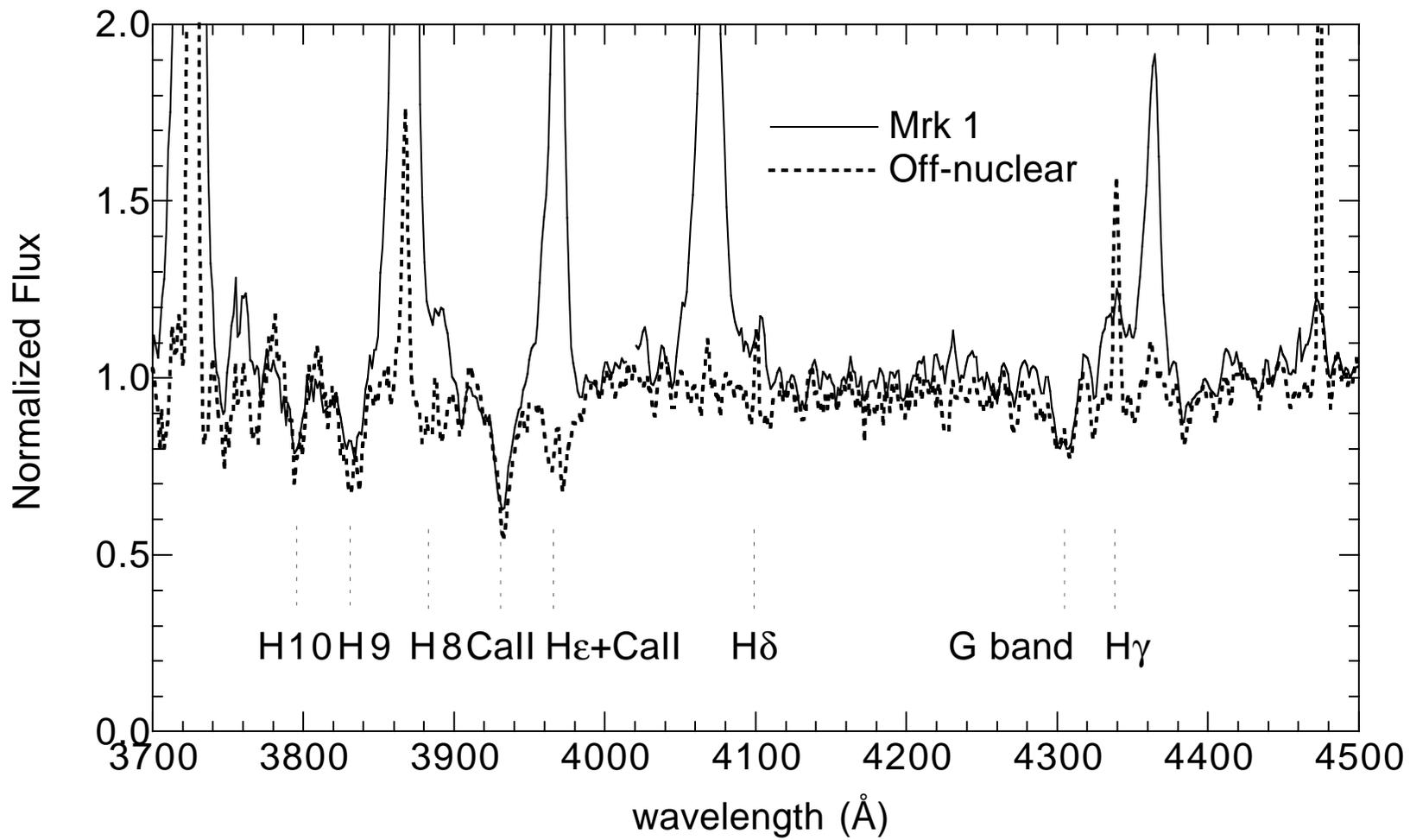
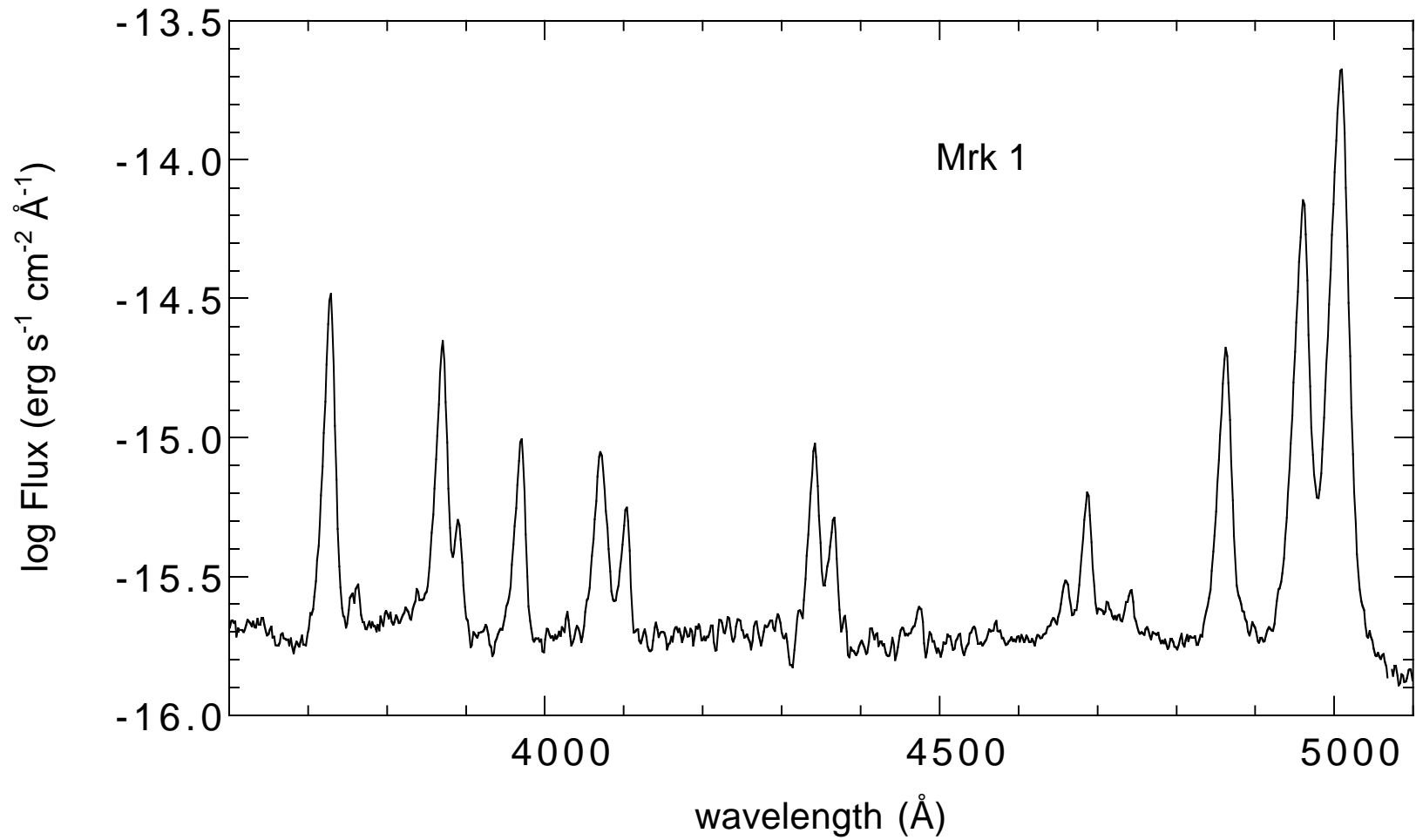
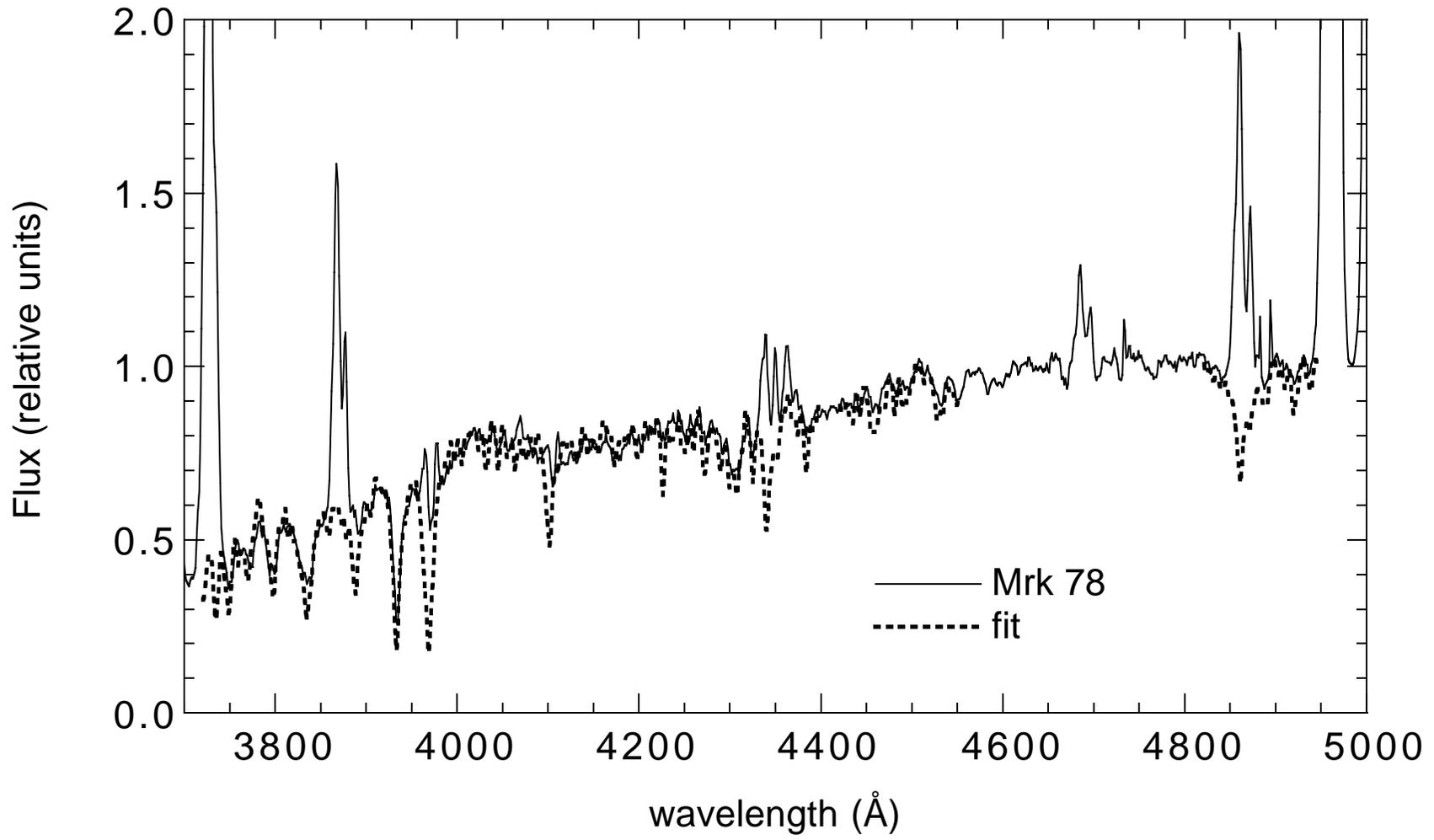
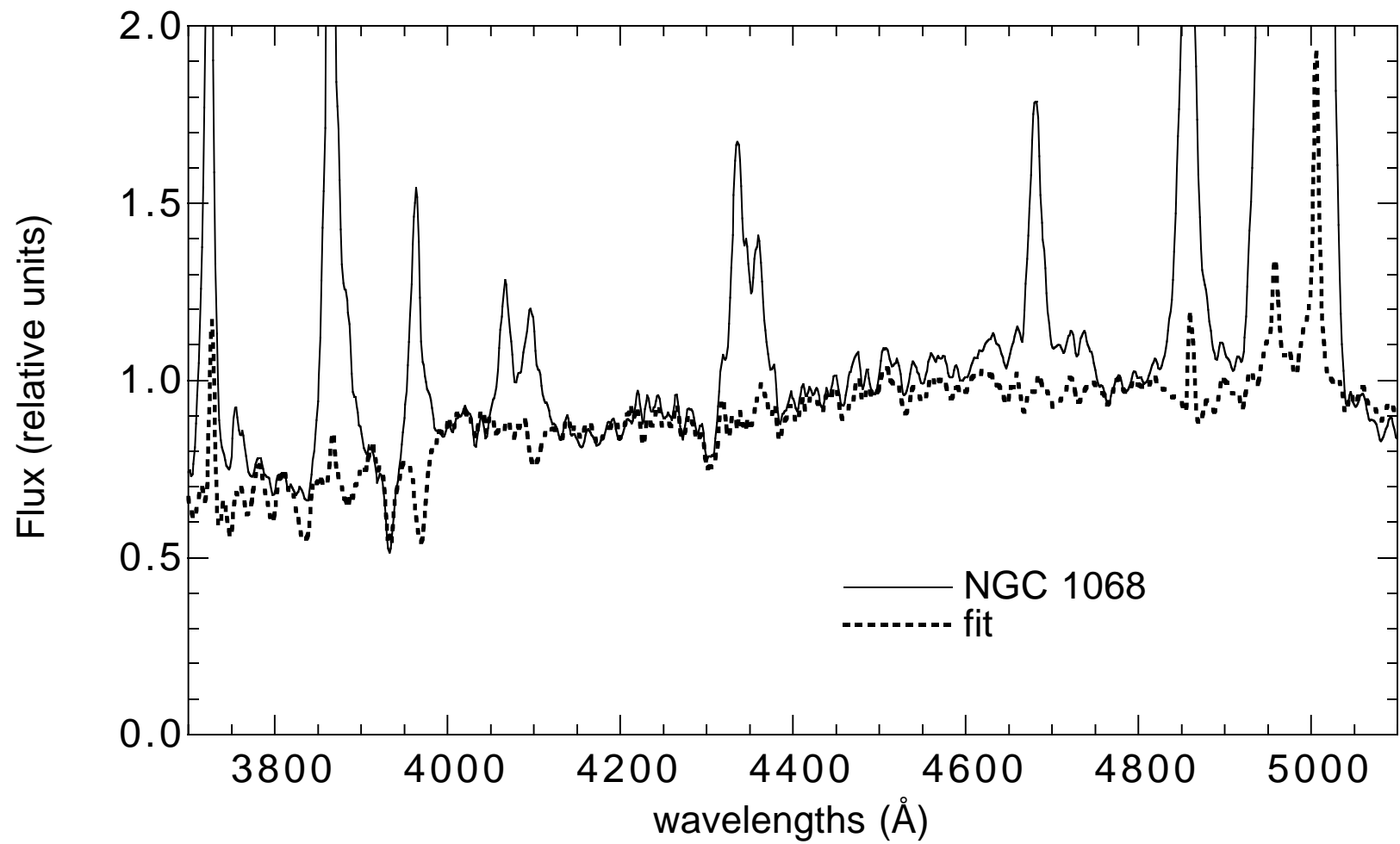
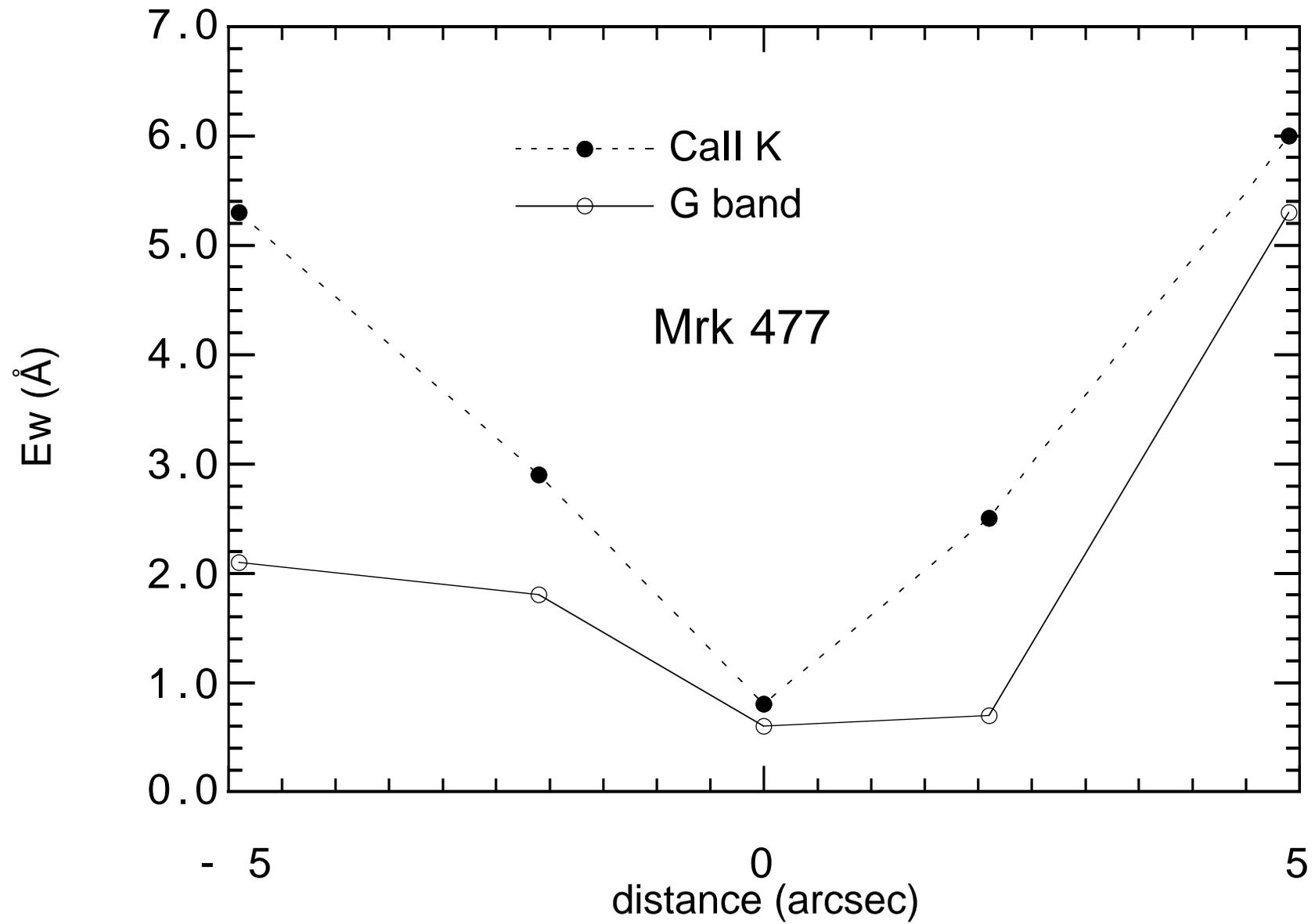
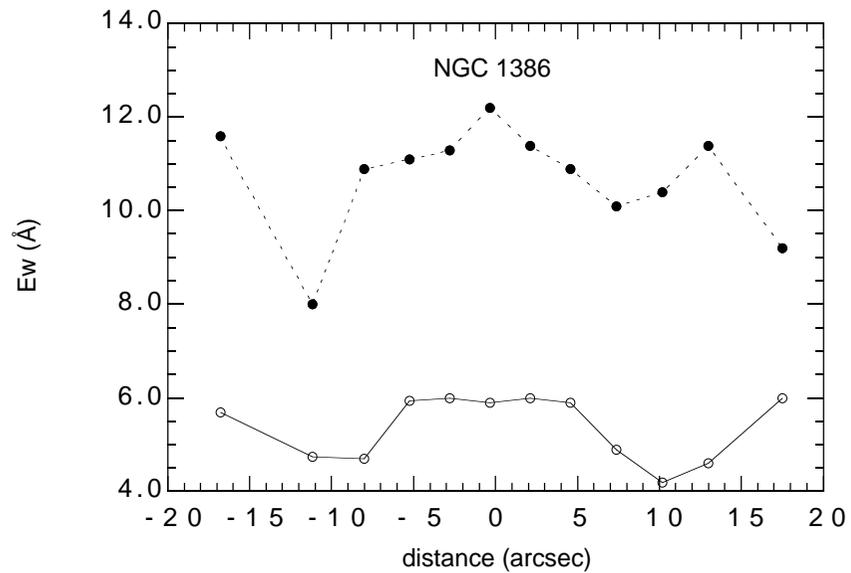
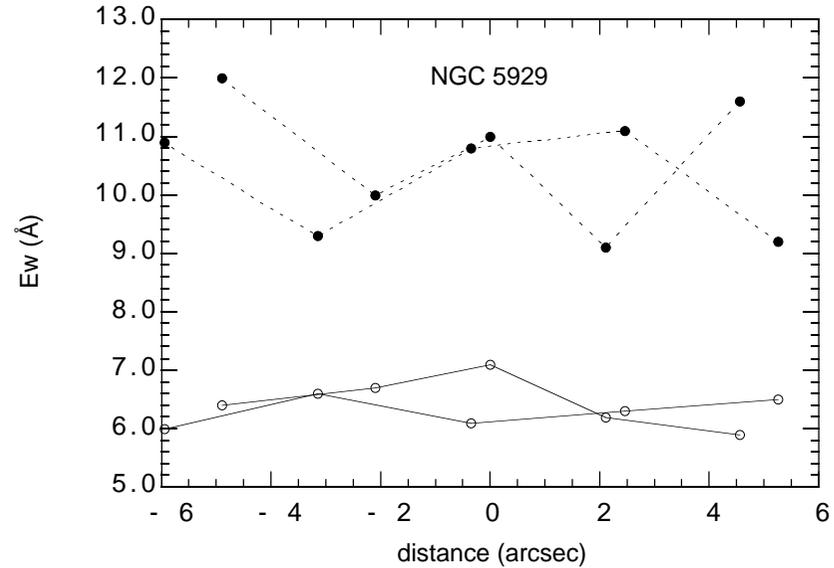
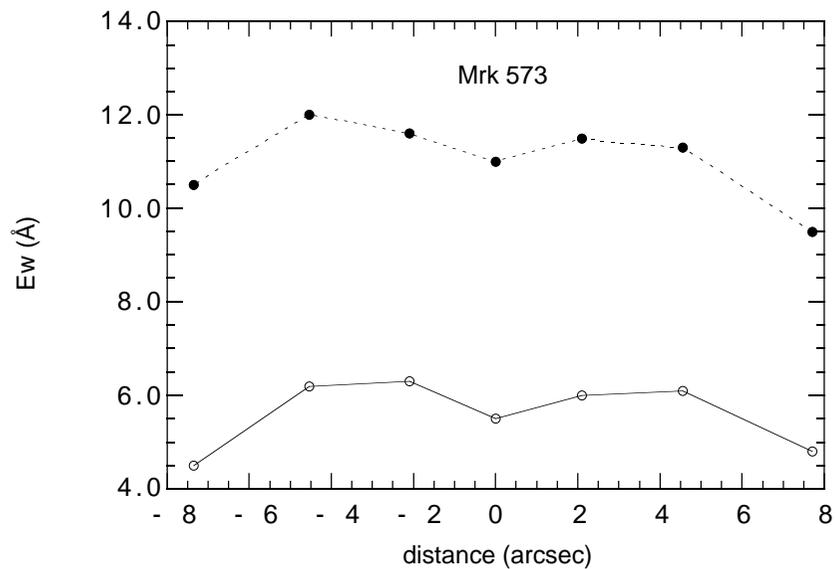
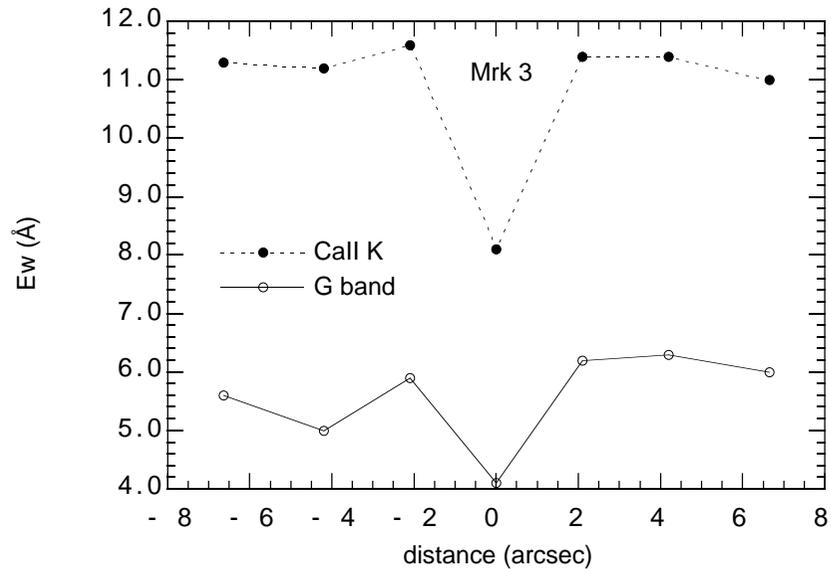
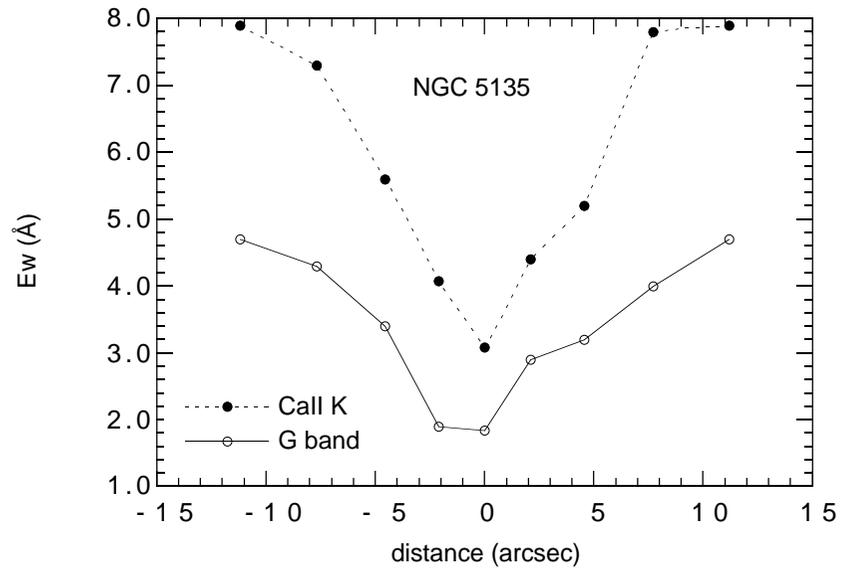
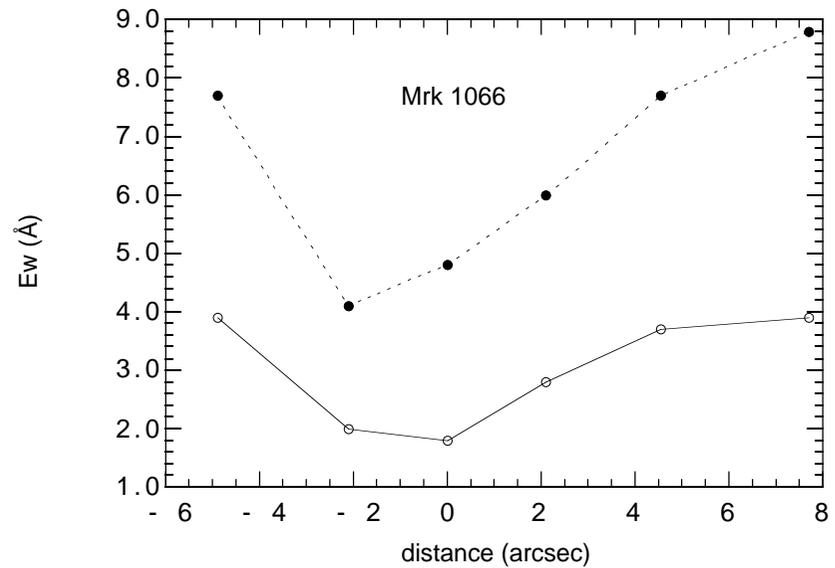
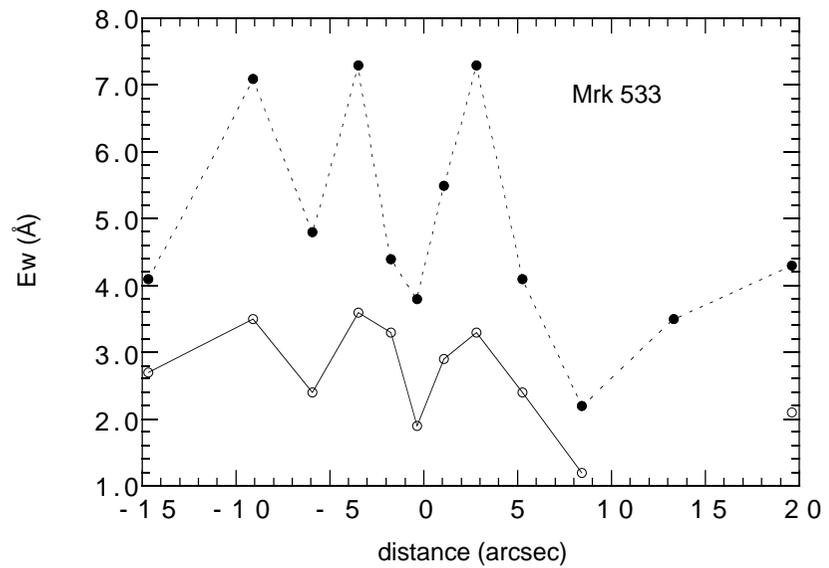
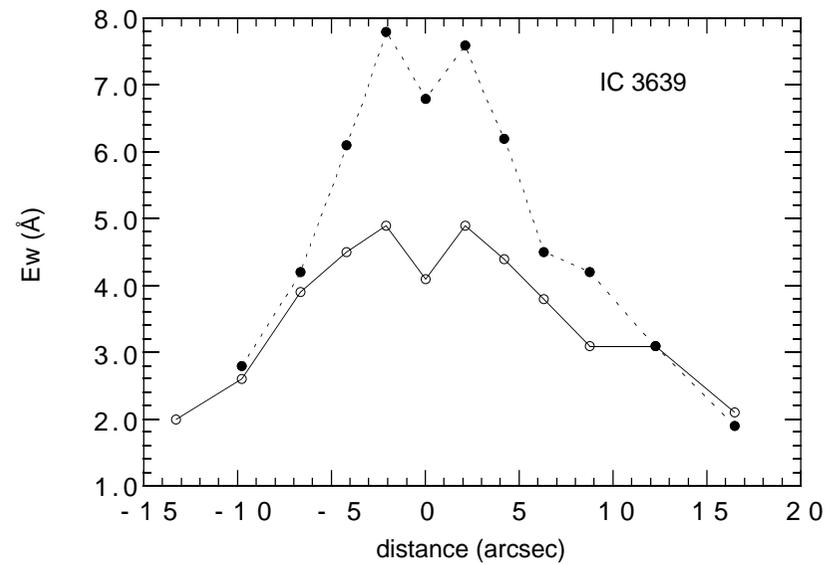
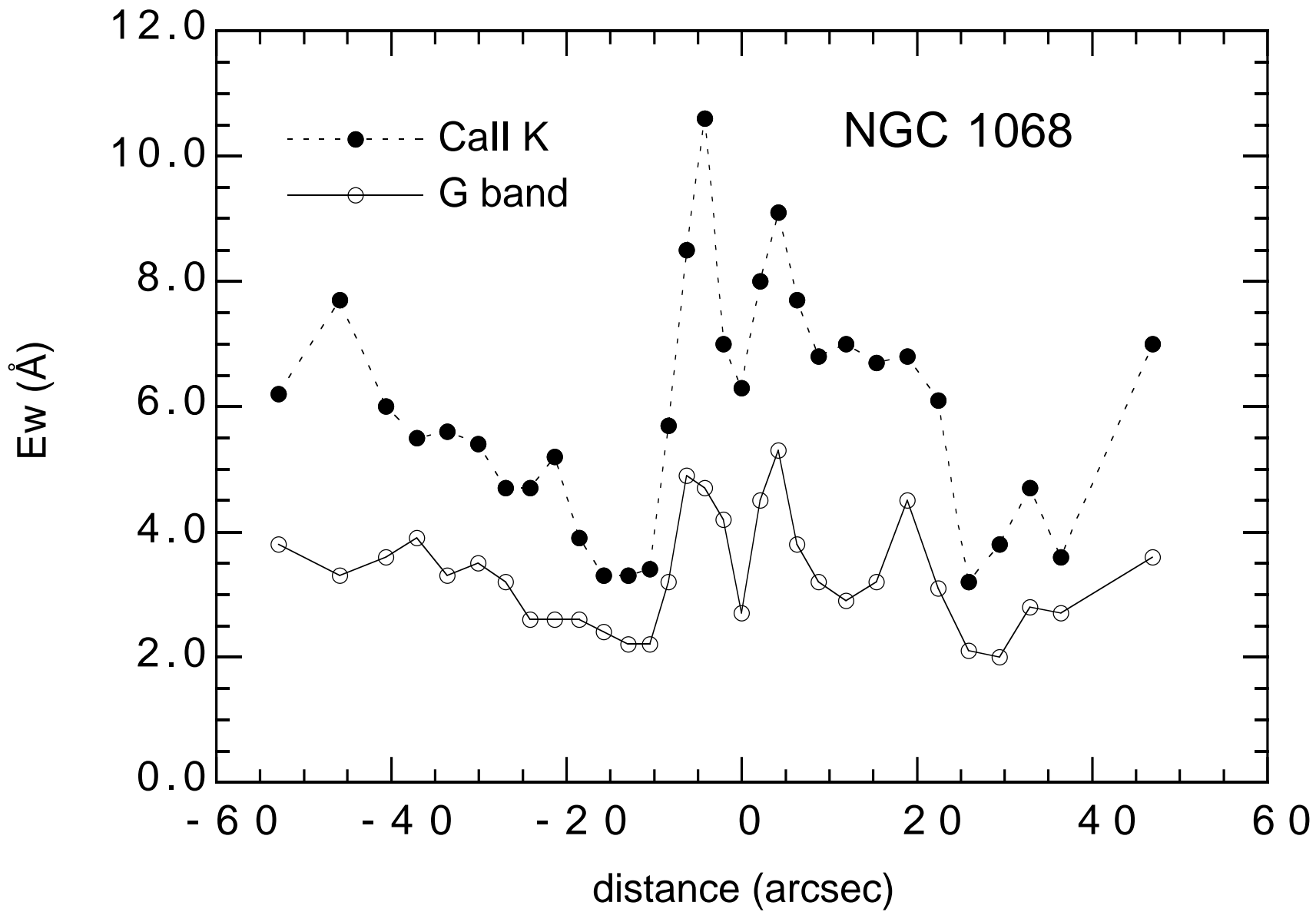
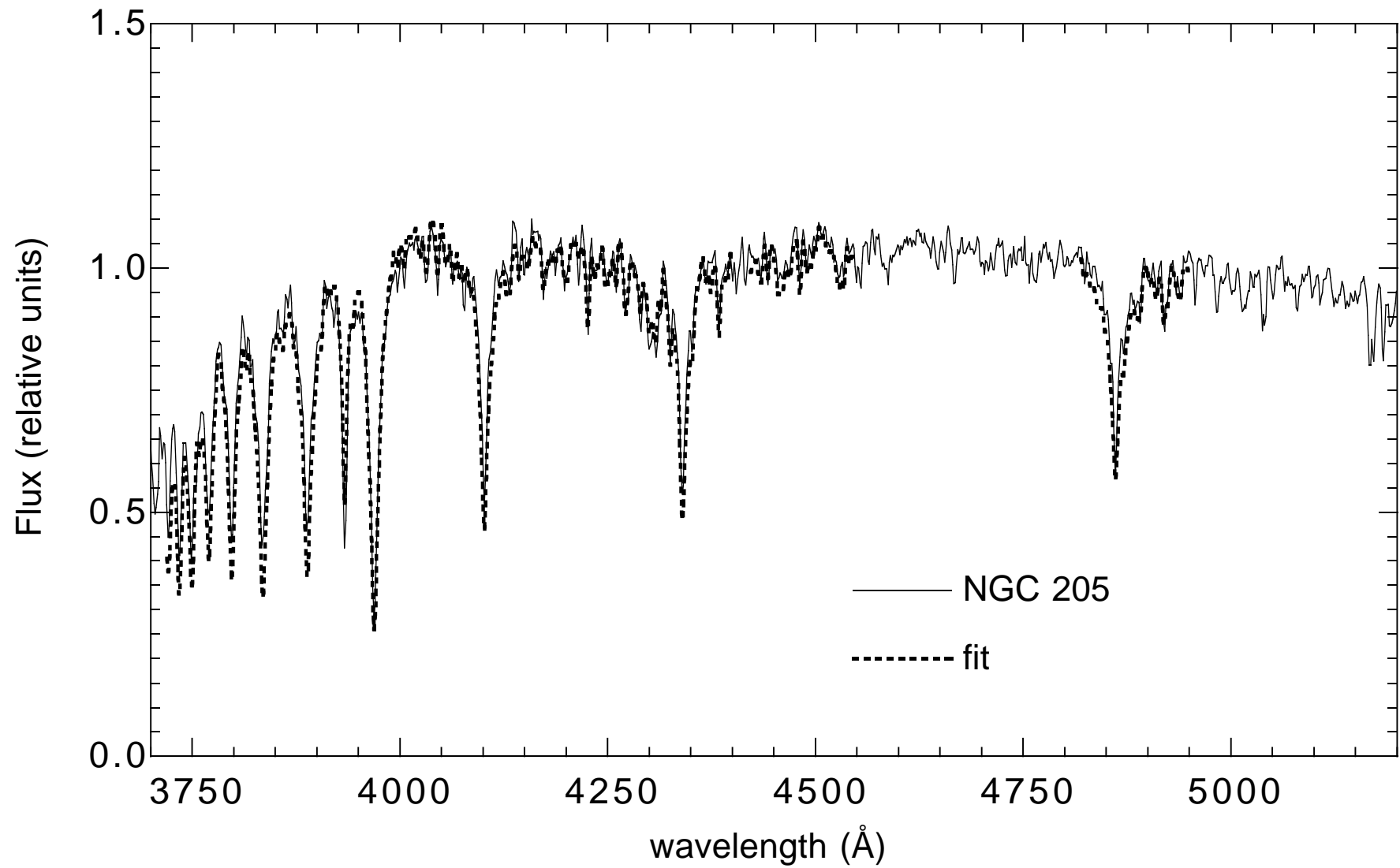
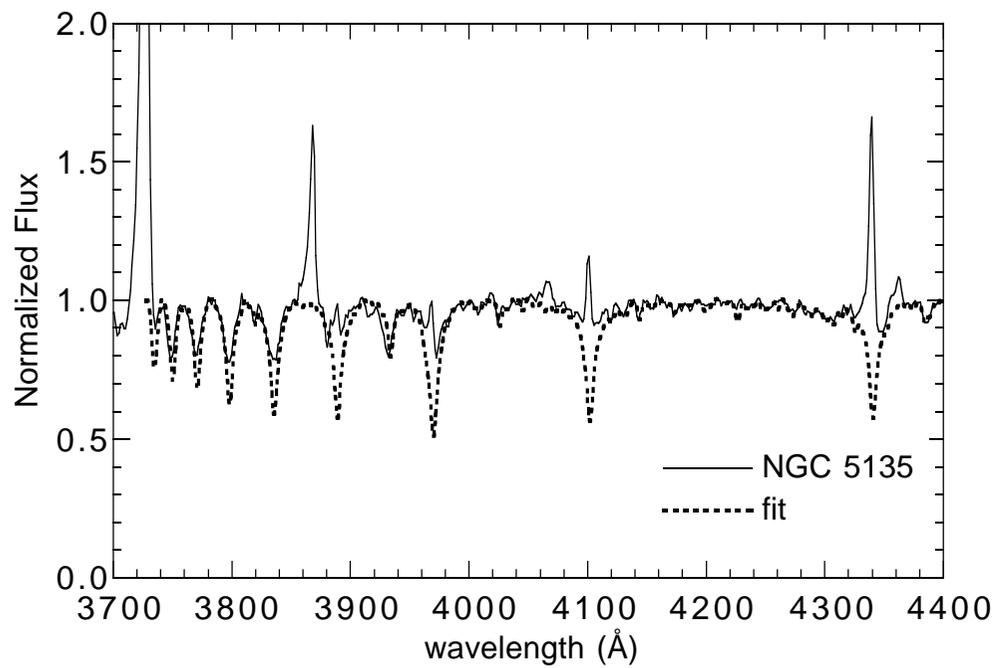
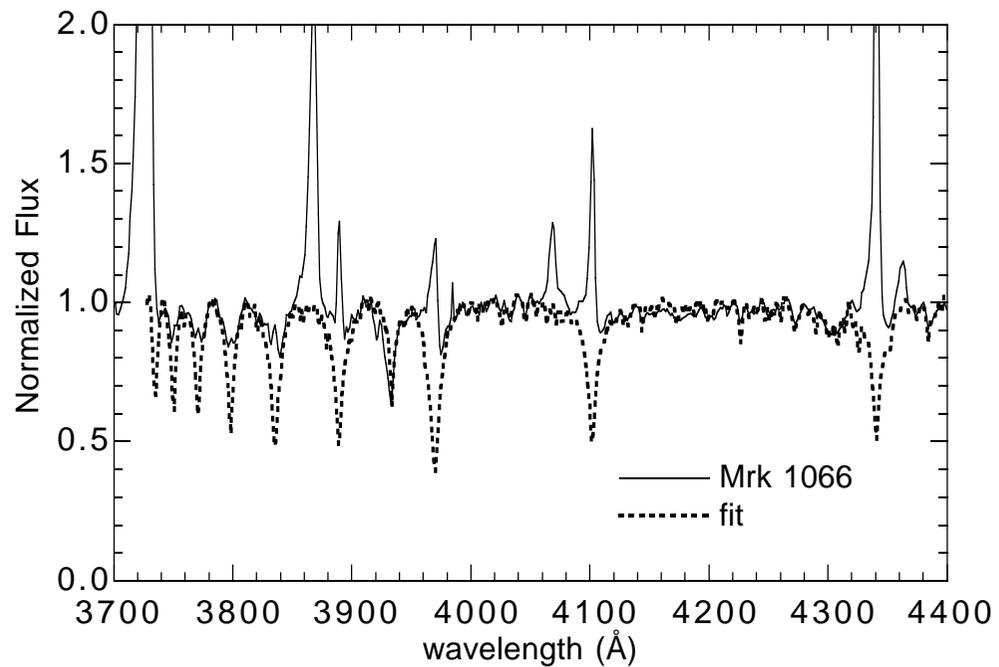
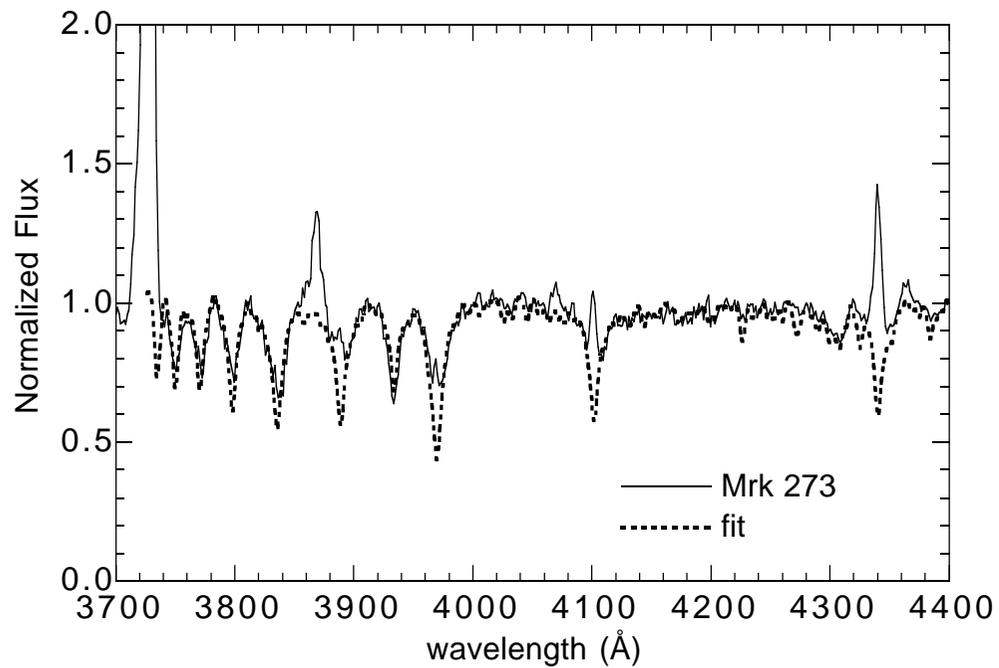
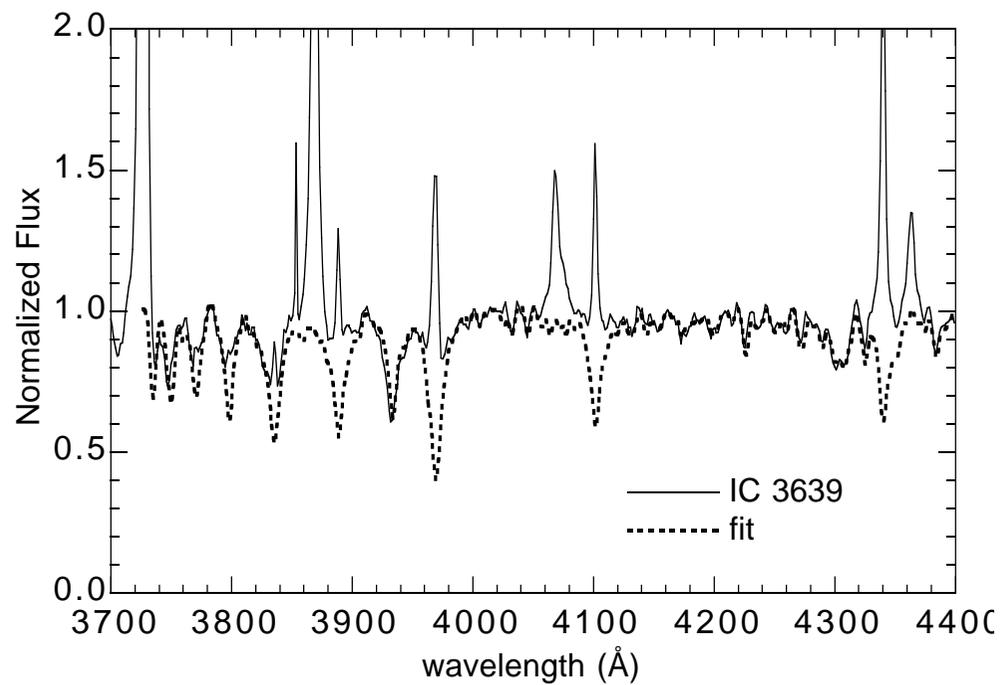
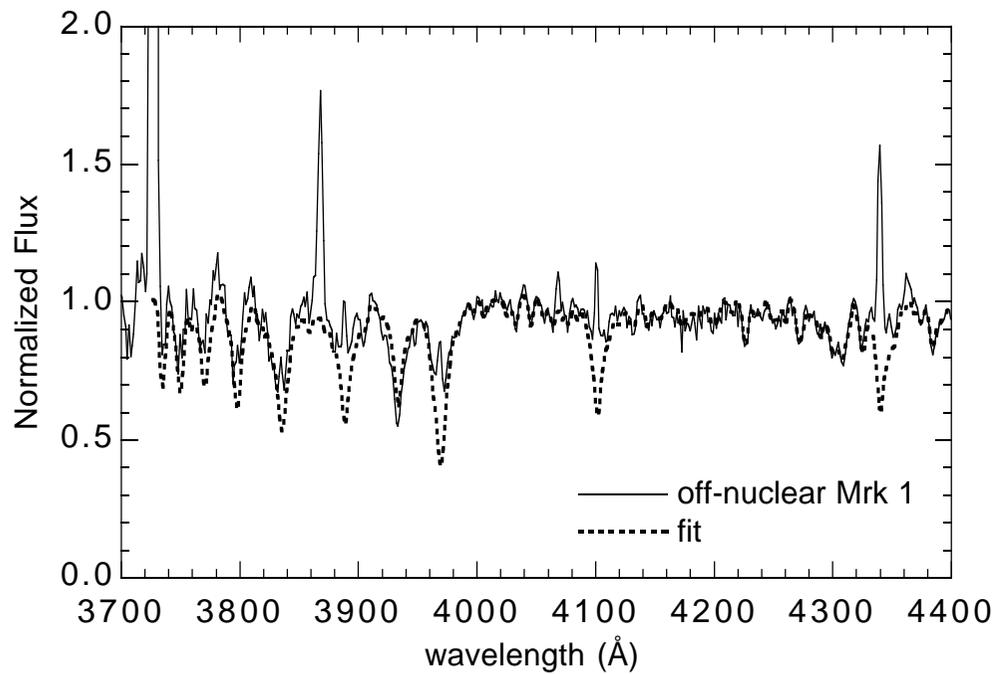
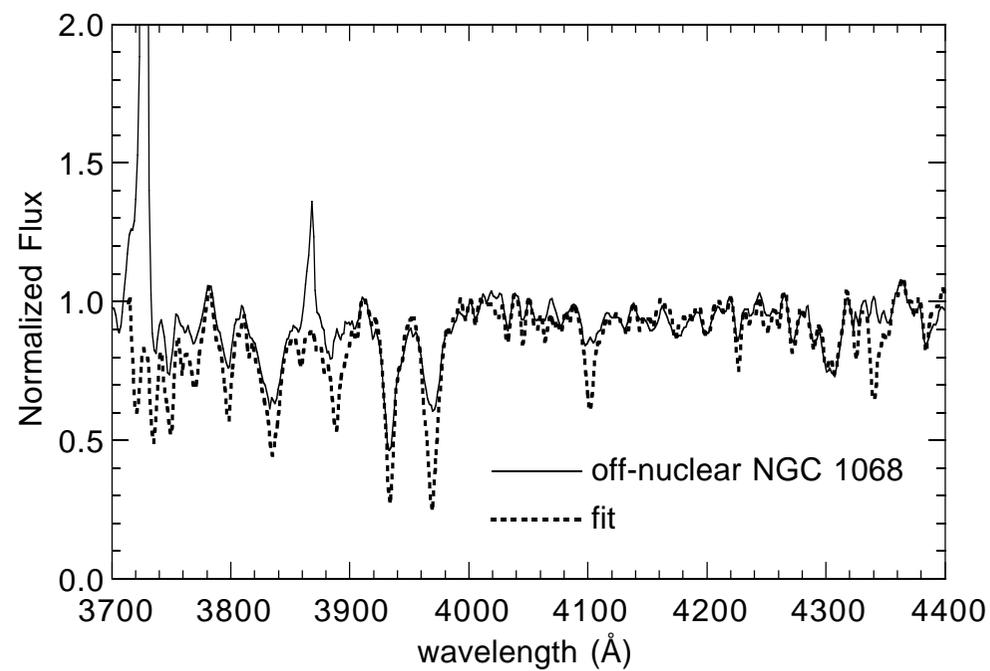
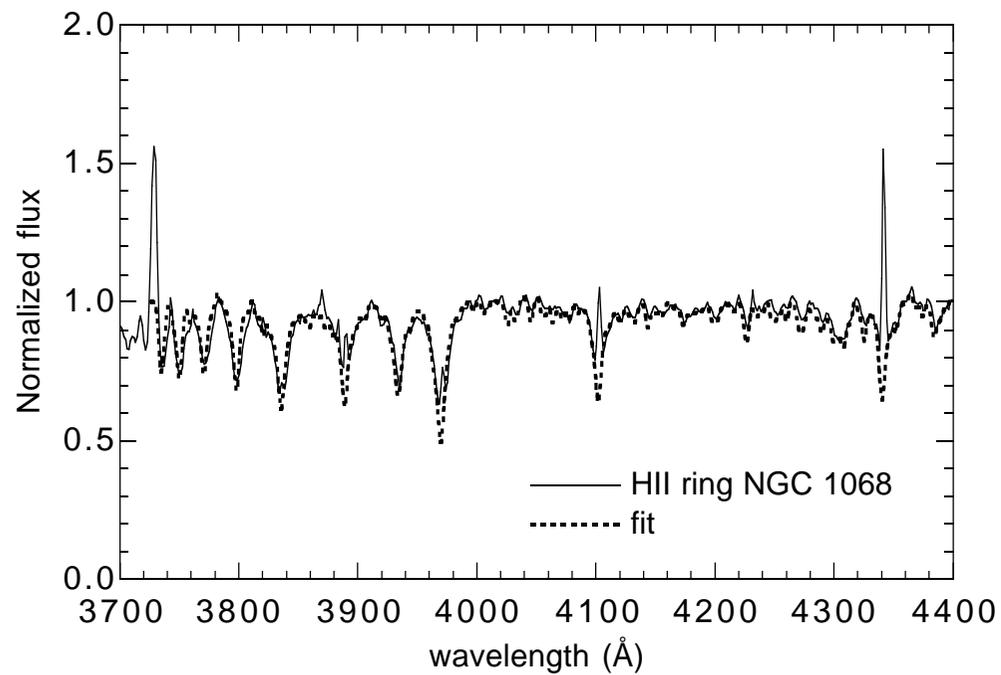
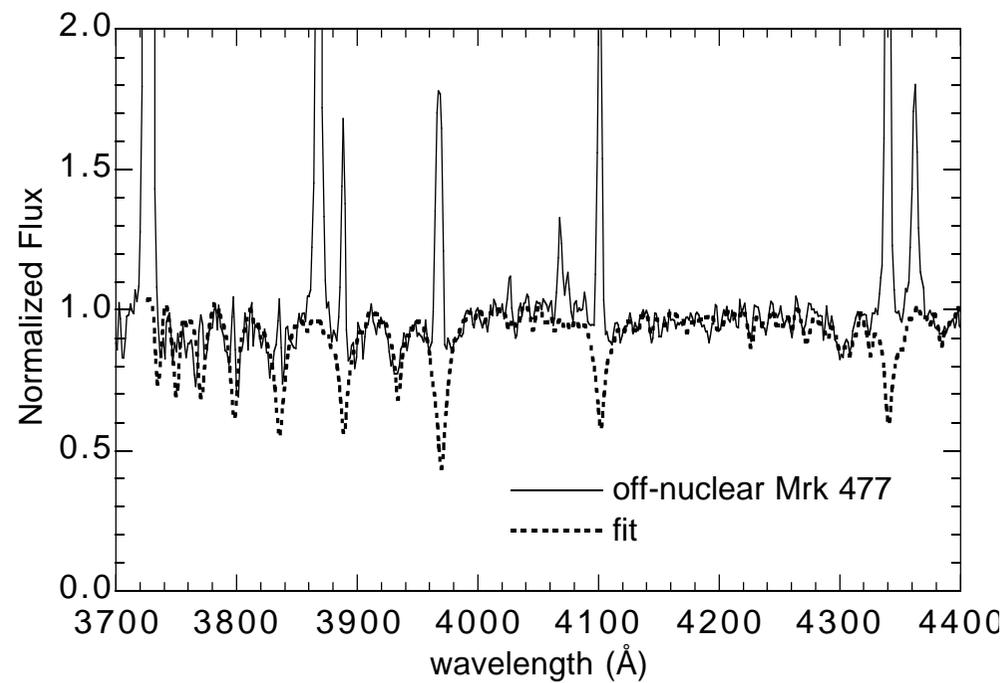
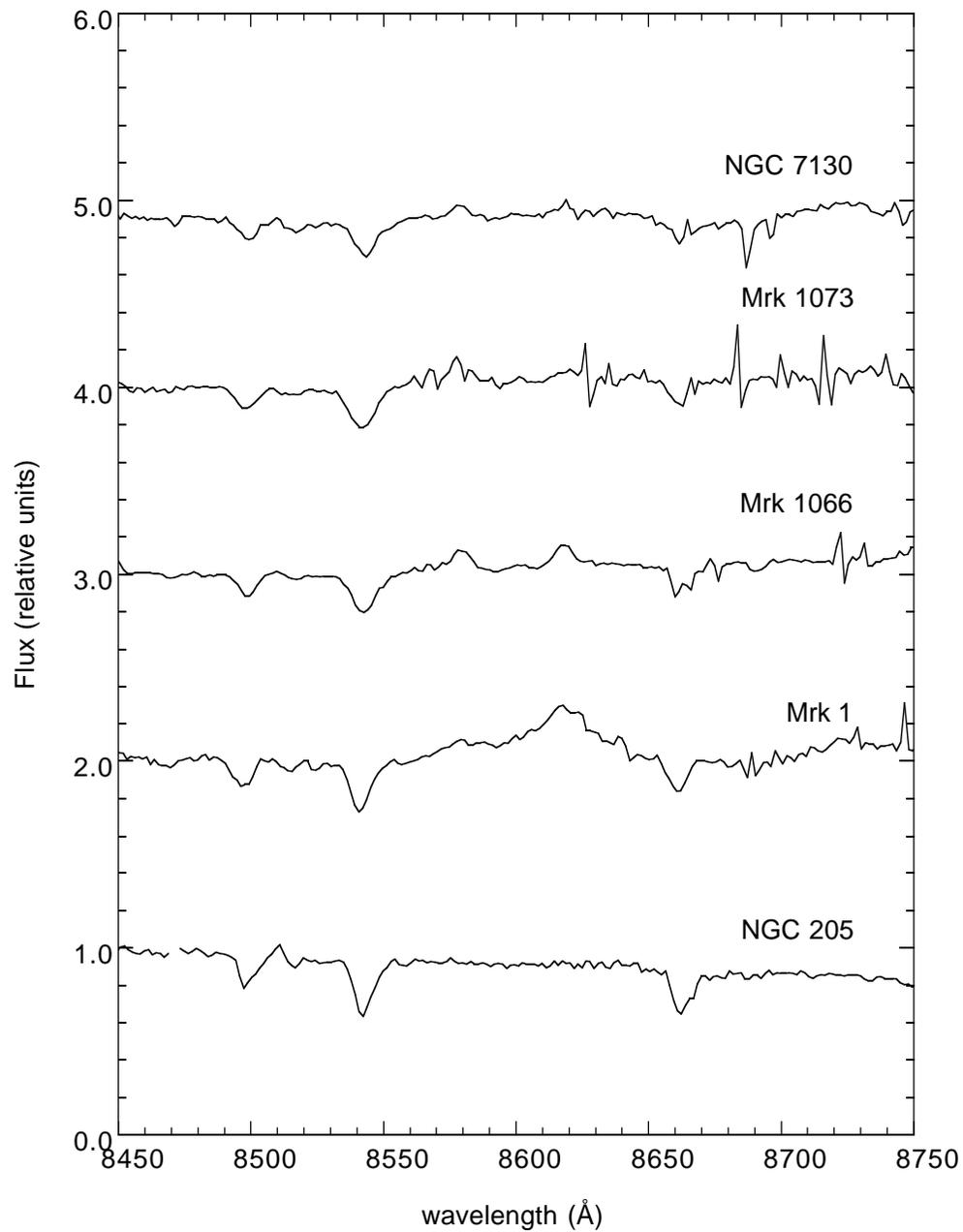
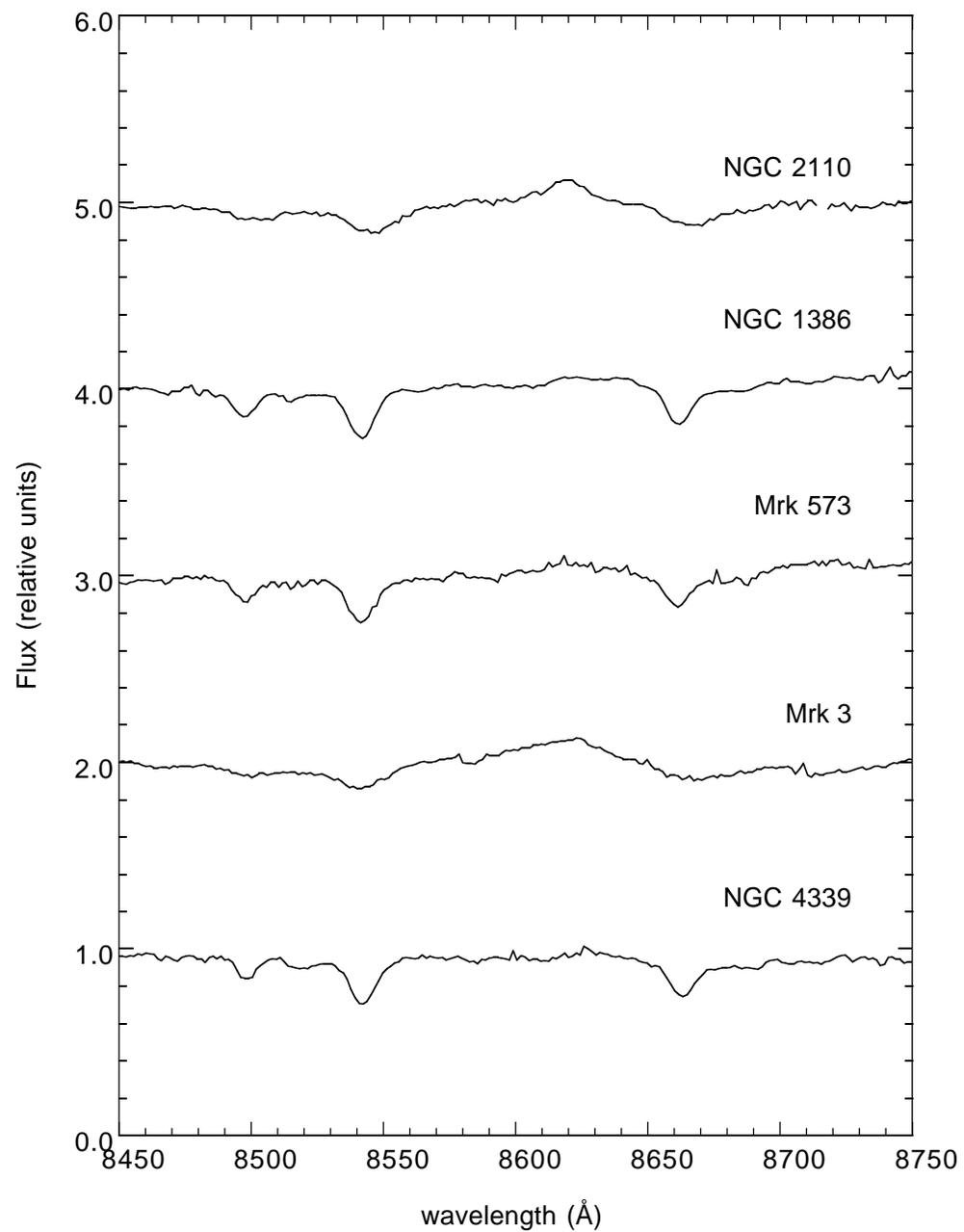
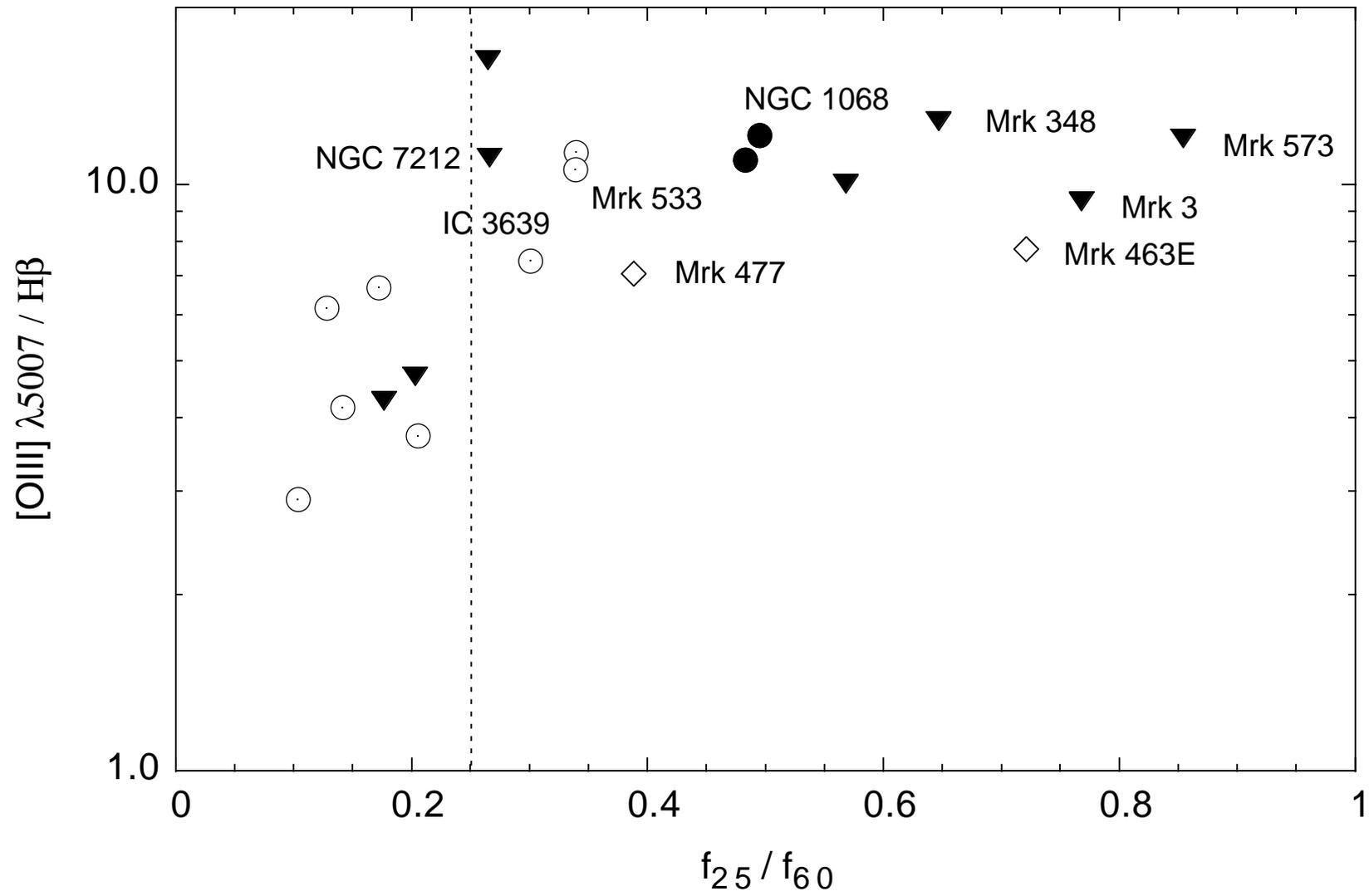
\end{document}